\documentclass[aps,pra,twocolumn]{revtex4}
\usepackage{graphics}
\usepackage[dvips]{epsfig}
\usepackage{subfigure}
\usepackage{multirow}
\usepackage{amssymb}
\usepackage{amsmath}
\usepackage{verbatim}

\begin{document}

\title{Inelastic Collisions and Chemical Reactions of Molecules at Ultracold Temperatures}



\author{Goulven Qu{\'e}m{\'e}ner, Naduvalath Balakrishnan}
\affiliation{Department of Chemistry, University of Nevada Las Vegas,
Las Vegas, NV 89154, USA}

\author{Alexander Dalgarno}
\affiliation{ITAMP, Harvard-Smithsonian Center for Astrophysics,
Cambridge, MA 02138, USA}

\date{\today}

\maketitle

\font\smallfont=cmr7

\tableofcontents{}

\section{ Introduction}

The development of techniques for cooling and trapping of a wide variety of atomic and molecular species 
in recent years has created exciting opportunities for probing and controlling atomic and molecular encounters with unprecedented
 precision~\cite{Doyle04}. While many of the initial studies of cold atoms and molecules were centered on the creation of dense samples
 of cold and ultracold matter, more recent work has focused on the manipulation and control of
intermolecular interactions, with the ultimate aim of achieving quantum control of atomic and molecular collisions~\cite{Roman-pccp}. Though the ideas of quantum control
 of chemical reactions were proposed many years ago, 
the ability to create ultracold  molecules in specific quantum states has given further 
stimulus to this field. Its development requires that molecular properties and collisional behavior be well understood at cold
 and ultracold temperatures where the dynamics of molecules
 are dramatically different compared to collisions at elevated
 temperatures. Over the last ten years significant progress has been achieved both in
 theoretical and 
experimental works. The experimental  methods such as photoassociation spectroscopy, 
magnetic tuning of Feshbach resonances, 
buffer-gas cooling, and Stark deceleration~\cite{Bahns00,Masnou01,Bethlem03,Hutson06} 
have been
 developed and applied to a variety of molecular systems. 
Novel methods 
to study ultracold chemical reactions involving ion - molecule systems in a linear Paul trap
have been proposed ~\cite{Willitsch08}. External control of
 chemical reactions using electric and magnetic fields is another area of active interest~\cite{roman-review}. The aim of this chapter 
is to provide an overview of recent progress in characterizing molecular processes and chemical reactions at cold and 
ultracold temperatures with particular emphasis on theoretical developments in 
quantum dynamics simulations  of atom - molecule collision systems over the last ten years.  \\

In contrast to scattering at thermal energies, ultracold collisions offer 
fascinating and unique opportunities to study molecular encounters in the extreme quantum regime where 
the entire collision can be dominated  by a single partial wave. 
 One of 
the main motivations of current experimental efforts to create dense samples of ultracold molecules is to study the possibility
of chemical reactions at 
temperatures close to absolute zero. While Wigner's law \cite{Wigner,bala97b}
 predicts that rate coefficients of exothermic processes are finite in the
 zero-energy limit it does not say if the rate coefficient will be large enough for reactions to be observable in an experiment.
 Nor does it say anything about how the rate coefficients at zero temperature depend on the interaction potential. Most 
chemical reactions between neutral atoms and molecules involve an energy barrier and it is not clear if chemical reactions between them
generally occur with measurable rates at ultracold temperatures. 
Calculations for the F + H$_2$ reaction which proceeds by tunneling at low energies 
have shown that the reaction may occur with a significant rate at ultracold temperatures~\cite{bala-cpl-2001}. There is also experimental and theoretical 
evidence that chemical reactions involving heavy-atom tunneling 
of carbon~\cite{zuev03} and fluorine~\cite{weck06} atoms can occur with significant rate coefficients at low 
temperatures. \\

Photoassociation experiments involving alkali-metal systems have
stimulated considerable 
interest in chemical
 reactivity in alkali-metal dimer - alkali-metal atom collisions at ultracold temperatures~\cite{Hutson07a}. 
Rearrangement collisions in identical particle alkali-metal trimer systems occur without energy barrier and
recent studies have 
demonstrated 
that chemical reactions in alkali-metal atom - alkali-metal dimer collisions may be very fast at 
ultracold temperatures.
Unlike tunneling-dominated reactions, the limiting values of the rate coefficients for alkali-metal systems 
are less sensitive to vibrational excitation of the 
dimer. \\

In this chapter we give an overview of recent studies of ultracold atom - molecule collisions focusing on non-reactive and reactive
 systems and the effect of vibrational excitation of the molecule on the collisional outcome. We will discuss both 
tunneling-dominated and barrierless reactions and examine recent efforts in extending these studies to ionic systems as well as
 molecule - molecule systems. We consider mostly the novel aspects of collisional dynamics of atom - diatom systems
 at cold and ultracold temperatures
 with illustrative results for specific systems. 
For more comprehensive discussion of cold and ultracold collisions including reactive and non-reactive
processes and the effect of external fields we refer the reader to 
several review articles~\cite{Hutson06,roman-review,weck06,Hutson07a,bodo-review}
that have appeared in the last few years.
For details of the theoretical formalisms we refer to the chapters by Hutson
and by Tscherbul and Krems.

\section{ Inelastic atom - molecule collisions}

Theoretical studies of ultracold molecules received intense interest after the success of 
ultracold atom photoassociation and buffer gas cooling 
experiments which demonstrated that a wide array of molecular systems with thermal and non-thermal vibrational energy distributions can be 
created in ultracold traps. The collisional loss of trapped molecules is an important issue in these experiments.
Photoassociation produces molecules in highly excited vibrational levels. Whether the excited molecules decay
 by vibrational quenching or through chemical reaction is an intriguing question. While an extensive literature exists on the 
collisional relaxation of vibrationally excited molecules at elevated temperatures 
not much was known on the  magnitude
 of the relaxation rate coefficients at temperatures lower than one Kelvin. Though a few earlier 
reports~\cite{takayanagi87,hancock89,takayanagi90} on atom - diatom collisions 
in the Wigner threshold regime had been published, a detailed investigation of the 
 dependence of the 
relaxation rate coefficients on the internal energy of molecules 
and their sensitivity to details of the interaction potential has not been carried out. 
Here, we give a 
brief account of recent quantum dynamics calculations of vibrational and rotational energy transfer in atom - diatom collisions at cold
and ultracold temperatures. We focus on a few representative systems to illustrate the main 
features of energy transfer in non-reactive 
atom - molecule collisions at ultracold temperatures and we show how the corresponding rate coefficients are influenced by rotational or vibrational excitation of the molecule.

\subsection{ Vibrational and rotational relaxation}

\subsubsection{Collisions at cold and ultracold temperatures}

As discussed in the chapter by Hutson, at very low energies, 
scattering is dominated by s-waves and the scattering cross section can be
expressed in terms of a single parameter called the scattering length. For single-channel
scattering where only elastic scattering is possible, the scattering length is a real quantity and the magnitude of the cross section in the s-wave limit
 is given by $\sigma=4\pi a^2$ where $a$ is the scattering length. 
For multichannel scattering, as in vibrationally or rotationally inelastic collisions of molecules, the 
scattering length is a complex number and it is denoted as 
$a_{vj}=\alpha_{vj}-i\beta_{vj}$ 
where $v$ and $j$ are, respectively,
 the initial vibrational and rotational quantum numbers of the molecule 
\cite{bala97b,bala98}. The limiting
value of the 
elastic cross section in the presence of inelastic scattering 
is given by $\sigma_{vj}^{el}=4\pi |a_{vj}|^2=4\pi(\alpha_{vj}^2+\beta_{vj}^2)$.
 The total inelastic quenching cross section from a given initial rovibrational
level of the molecule is related to the imaginary part of the scattering length 
through the relation 
$\sigma_{vj}^{in}=4\pi\beta_{vj}/\text{k}_{vj}$ where k$_{vj}$ is the wave vector in the 
incident channel. The quenching rate coefficient becomes constant at ultralow 
temperatures and is given by $k^{in}_{vj}=4\pi\hbar\beta_{vj}/\mu$  at zero temperature
where $\mu$ is the 
reduced mass of the collisional system. Thus, the rovibrational relaxation rate coefficients
attain finite values for different initial vibrational and rotational levels of the molecule. 
The dependence of the rate coefficients on $v$ and $j$ has been an important issue in cold
molecule research because exothermic vibrational and rotational relaxation 
collisions are
a major pathway for trap loss in cooling and trapping experiments.  \\

\begin{figure}[h]
\begin{center}
\includegraphics*[width=8cm,keepaspectratio=true,angle=0]{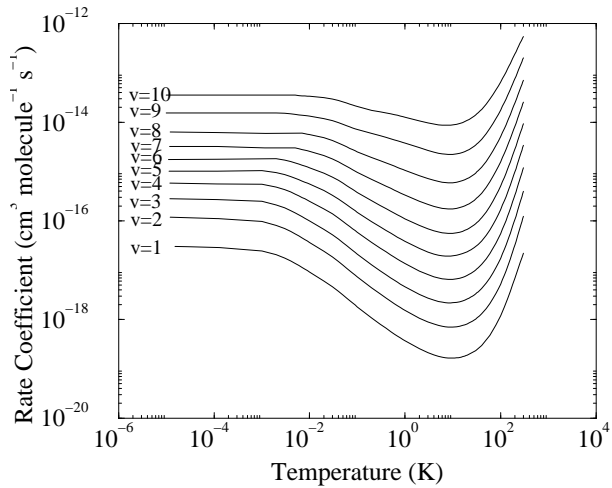} 
\caption{Rate coefficients for the quenching of H$_2(v,j=0)$ by collisions with He atoms as functions
of the temperature for $v=1-10$ of the H$_2$ molecule. 
Reproduced with permission from Balakrishnan et al.~\cite{bala98}.
\label{He-H2-rate}
}
\end{center}
\end{figure}

Initial studies of rotational and vibrational relaxation of atom - molecule systems at cold and ultracold temperatures have mostly
focused on van der Waals systems such as He$-$H$_2$~\cite{bala98,forrey98,forrey99a,forrey99b},  He$-$CO~\cite{bala00,zhu01}, and 
He$-$O$_2$~\cite{bala01b}. 
Owing to the importance of some of these systems in astrophysical environments, 
extensive calculations of low-temperature behavior of rate coefficients have been performed for 
collisions of
H$_2$~\cite{bala98} and CO~\cite{bala00,zhu01} with both $^3$He and $^4$He. For both systems reasonably accurate intermolecular potentials have been reported. The initial 
calculations on the He$-$H$_2$ system employed the potential energy surface (PES) of Muchnick and Russek (MR)~\cite{muchnik}. For He$-$H$_2$
 vibrational excitation of the H$_2$ molecule has a dramatic effect on the zero-temperature quenching rate coefficients. As
 illustrated in  Fig.~\ref{He-H2-rate}, the vibrational quenching rate coefficients 
increase by about three orders of magnitude between $v=1$ and 
$v=10$ of the H$_2$ molecule~\cite{bala98}.

The quenching rate coefficients exhibit a minimum at around 10~K which roughly corresponds to the
depth of the van der Waals interaction potential. This behavior appears to be a characteristic of vibrational
quenching rate coefficients. For incident energies lower than the well depth the rate 
coefficient exhibits a minimum and with subsequent decrease in temperature the rate coefficient begins to increase before attaining the 
Wigner limit. For systems with deeper van der Waals wells the minimum is shifted to higher temperatures. 
Measurements of vibrational relaxation rate coefficients for the H$_2-$CO system have confirmed this
behavior~\cite{Wilson93}.

Balakrishnan, Forrey, and 
Dalgarno~\cite{bala97a} also investigated vibrational relaxation of H$_2$ in collisions with H atoms for vibrational quantum numbers $v=1-12$ of the H$_2$ 
molecule. They adopted a non-reactive scattering formalism and neglected the rotational motion of the H$_2$ molecule. 
The calculations showed that vibrational relaxation rate coefficients  are  strongly dependent on the initial vibrational level of
the H$_2$ molecule. The relaxation rate coefficients were found to increase by about seven orders of magnitude between 
vibrational levels $v=1$ and $v=12$.
The dramatic variation in the rate coefficients with increase in vibrational excitation was explained 
in terms of the matrix elements of the interaction potential between the vibrational wavefunctions as functions of the  
atom - molecule center-of-mass separation.  \\

\begin{figure}[h]
\begin{center}
\includegraphics*[width=8cm,keepaspectratio=true,angle=0]{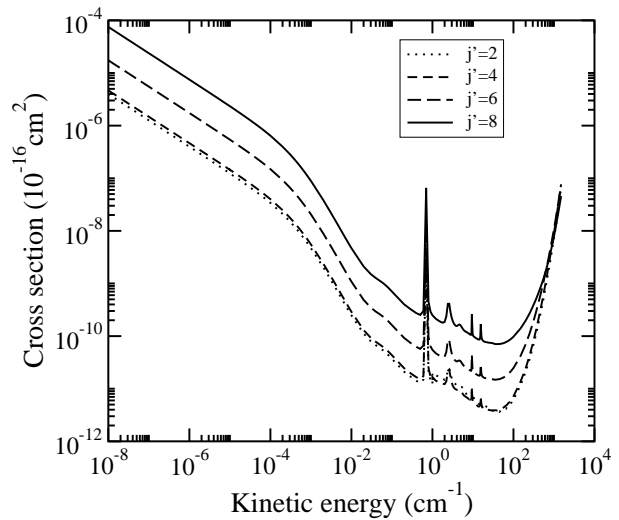}
\caption{ Cross sections for quenching of the $v=1,j=0$ level of D$_2$ in
collisions with Ar atoms resolved into the different rotational levels $j'$ in $v'=0$ as
 functions of the incident kinetic energy.
Reproduced with permission from Uudus et al.~\cite{Uudus05}. 
\label{Ar-D2-cross}
}
\end{center}
\end{figure}

Vibrational relaxation rate coefficients in atom - molecule systems are often influenced 
by van der Waals complexes formed during the collision process. Decay of these complexes leads 
to resonances in the energy dependence of the relaxation cross sections (see Fig. \ref{Ar-D2-cross} for Ar$-$D$_2$ collisions).  
Applying effective range theory
to describe ultracold collisions of
the He$-$H$_2$ system, Balakrishnan et al.~\cite{bala98} and Forrey et al.~\cite{forrey98} 
demonstrated that
 vibrational pre-dissociation lifetimes of resonances that lie close to the energy threshold
 can be derived accurately from the value of the zero-temperature quenching rate
coefficient. This formalism was extended to
 describe vibrational relaxation of trapped molecules and it was shown that the vibrational relaxation rate is controlled by the most 
weakly bound state of the van der Waals complex~\cite{forrey99b}. In a related work Dashevskaya et al.~\cite{dashev03} 
have shown that vibrational quenching of H$_2(v=1,j=0)$ at low temperatures can be described using a two channel approximation
within the quasi-classical method provided appropriate parameters are employed in the calculations. Subsequently, 
C{\^o}t{\'e} et al.~\cite{cote04} generalized this method to predict vibrational relaxation lifetimes of atom - diatom
 van der Waals complexes with energies near the dissociation threshold. \\

Unlike in thermal energy collisions, the presence of
a weakly bound state in the vicinity of a channel threshold can 
dramatically influence the cross sections in the ultracold regime. 
This is illustrated in Fig.~\ref{Ar-D2-cross} where 
the cross sections for vibrational relaxation of D$_2(v=1,j=0)$ in collisions with Ar
atoms~\cite{Uudus05} are presented over an energy range of $10^{-8}-10^3$ cm$^{-1}$. 
The cross sections exhibit a curvature characteristic
of a resonant enhancement in the energy range $10^{-5}-10^{-3}$ cm$^{-1}$. Such
enhancement of the cross section can occur when the interaction potential 
supports a virtual state or a very weakly bound state near the channel threshold 
leading to a  zero-energy resonance. The 
virtual state is characterized by a large negative scattering length while the bound state
is characterized by a large positive scattering length. For the present case the real
part of the scattering
length is large and positive ($\alpha_{10}=97.0$ \AA) and the resonance occurs from
the decay of a loosely bound van der Waals complex supported by
 the entrance channel potential. For energies below
10$^{-2}$~cm$^{-1}$ the cross section is dominated by s-wave scattering in the incident channel
and the zero-energy resonance arises from s-wave scattering in the entrance channel. \\

\begin{figure}[h]
\begin{center}
\includegraphics*[width=8cm,keepaspectratio=true,angle=0]{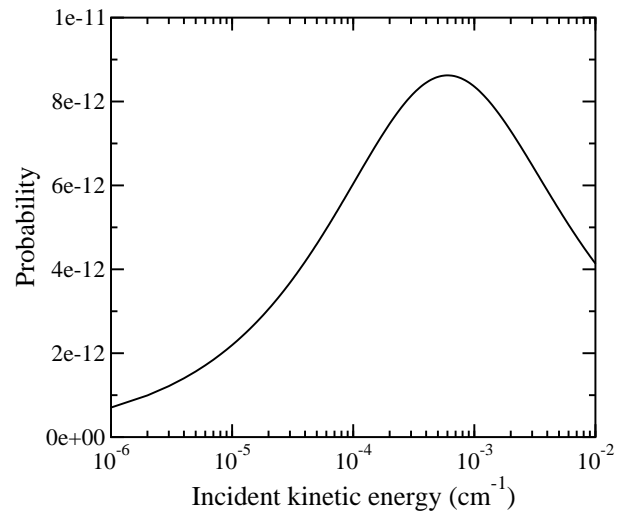} 
\caption{Total probability of quenching of the 
$v=1,j=0$ level of D$_2$ in collisions with Ar atoms as a function of the incident 
kinetic energy. The  peak value of the probability corresponds 
to a zero energy resonance. 
Reproduced  with permission from Uudus et al.~\cite{Uudus05}.
\label{Ar-D2-probability}
}
\end{center}
\end{figure}

The
resonant enhancement is more clearly seen in the plot of the reaction probability 
as a function of the kinetic energy 
shown in Fig.~\ref{Ar-D2-probability}.
The probability peaks at an energy of 
$6.0\times 10^{-4}$ cm$^{-1}$ which roughly corresponds to the binding energy of the
quasibound state. The resonance appears in the scattering 
calculations at energies above the threshold due to its close
proximity to the channel threshold. The binding energy of the quasibound state
can be estimated using the scattering length approximation \cite{bala97b,Uudus05}. 
The magnitude of the binding energy is given by 
$|E_b|=\hbar^2\cos{2\gamma_{10}}/(2\mu |a_{10}|^2)$
where $\mu$ is the reduced mass of the Ar$-$D$_2$ system,  $a_{10}=\alpha_{10}-i\beta_{10}$ is the scattering length for the $v=1,j=0$ level, and 
$\gamma_{10}=\tan^{-1}{(\beta_{10}/\alpha_{10})}$. This yields a
value of $|E_b|=4.9\times 10^{-4}$~cm$^{-1}$, in reasonable
agreement with the exact value derived from scattering calculations.

A more
accurate value of the binding energy can be obtained using the 
effective range formula given by Forrey et al. \cite{forrey98}:
\begin{eqnarray*}
|E_{b}|=\frac{\hbar^2}{\mu r_0^2}\left(1-\frac{\alpha_{10}r_0}{|a_{10}|^2}-
\sqrt{1-\frac{2\alpha_{10}r_0}{|a_{10}|^2}}\right)
\end{eqnarray*}
where $r_0$ is the effective range of the potential which may be evaluated by fitting the
low energy behavior of the phase shift for the elastic channel to the standard effective
range formula, $k_{10}\cot{\delta_{10}}=-1/\alpha_{10}+r_0k_{10}^2/2$. For Ar$-$D$_2(v=1,j=0)$
collisions, the effective range formula yields $r_0=16.32$ \AA. The resonance position
calculated using the effective range approximation is
$|E_b|=5.95\times 10^{-4}$ cm$^{-1}$, in excellent agreement with 
the value of $6.0\times 10^{-4}$ cm$^{-1}$ obtained from the scattering calculations. 
It is generally very difficult to accurately evaluate energies of such weakly
bound states using standard  bound state codes and the effective range formula provides 
a convenient and reliable method to calculate binding energies of weakly bound states
that lead to zero-energy resonances. \\

One of the challenging aspects of cold and ultracold collisions is the sensitivity to details of the interaction 
potential. Even the best available methods 
for the electronic structure calculations of PESs result in errors much 
larger than the collision energies in the cold and ultracold regime
 and the dynamics calculations are often sensitive to small changes in the interaction potential.
To explore the sensitivity of cold and ultracold collisions to details of the interaction potential, Lee et al.~\cite{lee05} 
performed a 
comparative study of the ultracold collision dynamics of the He$-$H$_2$ system using the MR
 potential and a more recent ab initio potential developed by Boothroyd, Martin and Peterson (BMP)~\cite{boothroyd03}. The
 BMP potential was considered to be an improvement, approaching chemical accuracy, over all conformations 
compared to the MR potential. However, significant differences were observed for vibrational 
relaxation of the $v=1,j=0$ state of the H$_2$ molecule in collisions with He computed using the two 
surfaces. The limiting value of the quenching rate coefficient on the BMP surface was 
found to be about three orders of magnitude larger than that of the MR surface. The difference was 
attributed to the more anisotropic nature of the BMP surface leading to larger values of the off-diagonal
 elements responsible for driving vibrational transitions. Indeed, it was found that the vibrational 
quenching of the $v=1,j=0$ level was dominated by the transition to the $v'=0,j'=8$ level which is driven by 
the high-order anisotropic terms of the interaction potential.   \\


To explore the behavior of inelastic collisions involving polar molecules at 
ultracold temperatures, Balakrishnan, 
Forrey and Dalgarno~\cite{bala00} investigated vibrational and rotational relaxation
of CO in $^4$He$-$CO collisions.
The dynamics of the He$-$CO system was found to exhibit significantly different features at low temperatures
 compared to the He$-$H$_2$ system. Quantum scattering calculations of $^4$He and 
$^3$He collisions with the CO molecule 
revealed that the larger reduced mass and the deeper van der Waals interaction potential 
of the He$-$CO system give rise to a number of shape resonances in the energy dependence of vibrational relaxation cross sections \cite{bala00,zhu01}. The effect of shape resonances 
on low temperature vibrational relaxation rate coefficients will be discussed
in the next subsection. 
The computed values of vibrational relaxation rate coefficients for both $^3$He and 
$^4$He collisions with CO($v=1$) have been found to be in good agreement with experimental data of Reid et
 al.~\cite{reid97} in the temperature range $35-100$~K. 
Calculations of vibrational relaxation rate coefficients for the He$-$CO system
in the temperature range $35-1500$~K have also been reported by Krems~\cite{Krems02a}.
He has shown that inclusion of the centrifugal distortion
of the vibrational wavefunction enhances the relaxation process, and that  the quenching
rate coefficients are sensitive to high-order anisotropic terms in the 
angular expansion of the interaction potential~\cite{Krems02b}.

Bodo, Gianturco and Dalgarno~\cite{bodo02-cpl} have 
extended the work of Balakrishnan et al.~\cite{bala00}
to study the vibrational relaxation
of excited CO($v=2,j=0,1$) molecules in collision with $^4$He atoms at ultra-low energies. 
They found that vibrational quenching 
of CO($v=2,j=0,1$) in collisions with $^4$He
is dominated by the $v=2 \to v=1$ transition.
The cross sections for the $v=2 \to v=0$ transition
were found to be about four orders of magnitude smaller than the single
quantum transitions for both $j=0$ and $j=1$ initial rotational levels. \\

Ultracold vibrational relaxation of a number of other molecules in collisions
with He atoms has been reported by a number of other investigators in recent years.
Stoecklin, Voronin, and Rayez~\cite{stoeklin03a} reported the vibrational
relaxation of F$_2$ in collisions with $^3$He atoms. In this study, they constructed the PES of He$-$F$_2$
using ab initio points obtained by high-level molecular electronic-structure calculations,
and reported the cross sections for elastic scattering and inelastic relaxation
of F$_2(v=0,1,j=0)$ for collision 
energies in the range $10^{-6} - 2000$~cm$^{-1}$.
A similar study has been reported by the same authors
for the $^3$He + HF$(v=0,1,j=0,1)$ system~\cite{stoeklin03b}. 
The vibrational quenching cross sections were
found to be very small compared to pure rotational
quenching, in agreement with the results for the He$-$CO system. This is due to
the weak dependence of the He$-$HF PES 
on the HF internuclear distance and the strong
anisotropy of the interaction potential.

Bodo and Gianturco~\cite{bodo03-jpc} presented a comparative study of
vibrational relaxation of CO$(v=1,2,j=0)$, HF$(v=1,2,j=0)$ 
and LiH$(v=1,2,j=0)$ in collisions with $^3$He and $^4$He atoms  
in the Wigner regime. The  quenching rate coefficients
were found to depend strongly on the collision partner.
They reported rate coefficients in the range $10^{-21} - 10^{-19}$~cm$^{3}$~s$^{-1}$ for
CO, $10^{-16} - 10^{-15}$~cm$^{3}$~s$^{-1}$ for
HF, and  $10^{-14} - 10^{-11}$~cm$^{3}$~s$^{-1}$ for LiH. 
The differences were attributed 
to the features of the 
intermolecular forces between the diatomic molecules and the He atoms.
The interaction potential of the He$-$CO system is almost isotropic and is characterized by
small vibrational couplings elements. 
The He$-$HF system is more anisotropic and the couplings between vibrational states 
are more significant.
The interaction potential 
of the He$-$LiH system is very anisotropic
and it exhibits strong vibrational couplings. \\

There is considerable ongoing experimental interest
in cooling and trapping  NH~\cite{Campbell07,Hoekstra07} 
and OH~\cite{Meerakker05a,Sawyer07} molecules using the buffer gas cooling and
Stark deceleration methods described in the chapters by Doyle and by Meijer. 
Krems et al.~\cite{krems03b} and Cybulski et al.~\cite{Cybulski05}
reported cross sections and rate coefficients
for elastic scattering and Zeeman relaxation in $^3$He$-$NH collisions
from ultralow energies to 10~cm$^{-1}$. The calculations were 
performed using the rigid rotor approximation and an accurate He$-$NH PES. 
It was demonstrated that the elastic scattering of NH molecules with He atoms
in weak magnetic fields is at least five orders of magnitude
faster than the Zeeman relaxation, which suggests that the NH molecule
is a good candidate for buffer-gas cooling. In a related study
Gonz{\'a}lez--S{\'a}nchez et al.~\cite{Gonzalez06b}
examined  rotational relaxation and spin-flipping in collisions of OH with 
He atoms at ultralow energies. 
They found that the rotational relaxation processes dominate the elastic process
as the collision energy is decreased to zero. \\

While theoretical prediction of rate coefficients for vibrational and rotational 
relaxation in a number of atom - diatom systems
has been made, comparable experimental results are not available for 
majority of these systems.
The first measurements of vibrational relaxation of molecules at temperatures below 1~K were reported 
 by Weinstein et al.~\cite{weinstein1998}. In their study, CaH molecules slowed down by elastic 
collisions with $^3$He buffer gas atoms were 
trapped in an inhomogeneous magnetic field. An upper bound of the rate coefficients for spin-flipping transitions
 in CaH as well as 
vibrational relaxation of CaH molecules in the $v=1$ vibrational level in collisions with $^3$He atoms
 were estimated at a temperature of about 500~mK. Balakrishnan 
et al.~\cite{bala03b} 
presented a theoretical analysis of the vibrational relaxation of CaH in 
collisions with $^3$He atoms based on quantum close-coupling calculations and 
an ab initio PES for the He$-$CaH system developed by Groenenboom and 
Balakrishnan~\cite{groenenboom03}. 
In a related study, 
Krems et al.~\cite{krems03} reported cross sections for spin-flipping transitions in CaH induced by collisions with $^3$He and 
obtained results in close agreement with the experimentally derived values
 of Weinstein et al~\cite{weinstein1998}. 
Krems et al. demonstrated that at low energies, spin-flipping transitions in the $N=0$ rotational level of
 $^2\Sigma$ molecules induced by structureless atoms occur through coupling to the rotationally excited $N>0$ levels and that the 
corresponding rate coefficients are determined by the 
spin-rotation interaction  with the transiently rotationally excited molecule. \\

Table~\ref{TAB1} provides a compilation of zero-temperature quenching rate coefficients for 
vibrational and rotational relaxation in a number of atom - diatom systems.

\begin{table}[h]
\begin{center}
\begin{tabular}{c c c c}
\hline
system & initial $(v,j)$ & $k_{T=0}$ (cm$^3$s$^{-1}$) & Ref.  \\ [0.5ex]
\hline
H + H$_2$ & $(v=1,j=0)$ & 1.0 $\times$ 10$^{-17}$ & \cite{bala97a} \\
\hline
$^3$He + H$_2$ & $(v=1,j=0)$ & 3 $\times$ 10$^{-17}$ & \cite{bala98} \\
               & $(v=10,j=0)$ & 3.6 $\times$ 10$^{-14}$ & \cite{bala98} \\
\hline
$^4$He + CO & $(v=1,j=0)$ & 6.5 $\times$ 10$^{-21}$ & \cite{bala00} \\
            & $(v=1,j=1)$ & 9.0 $\times$ 10$^{-19}$ & \cite{bala00} \\
\hline
$^3$He + CO & $(v=1,j=0)$ & 1.3 $\times$ 10$^{-19}$ & \cite{bodo03-jpc} \\
            & $(v=2,j=0)$ & 2.1 $\times$ 10$^{-19}$ & \cite{bodo03-jpc} \\
$^4$He + CO & $(v=1,j=0)$ & 5.3 $\times$ 10$^{-21}$ & \cite{bodo03-jpc} \\
            & $(v=2,j=0)$ & 1.3 $\times$ 10$^{-20}$ & \cite{bodo03-jpc} \\
\hline
$^3$He + CaH & $(v=0,j=1)$ & 3.5 $\times$ 10$^{-12}$ & \cite{bala03b} \\
             & $(v=1,j=0)$ & 2.6 $\times$ 10$^{-17}$ & \cite{bala03b} \\
\hline
$^3$He + HF & $(v=1,j=0)$ & 3.1 $\times$ 10$^{-16}$ & \cite{bodo03-jpc} \\
            & $(v=2,j=0)$ & 2.6 $\times$ 10$^{-15}$ & \cite{bodo03-jpc} \\
$^4$He + HF & $(v=1,j=0)$ & 8.1 $\times$ 10$^{-16}$ & \cite{bodo03-jpc} \\
            & $(v=2,j=0)$ & 6.5 $\times$ 10$^{-15}$ & \cite{bodo03-jpc} \\
\hline
$^3$He + LiH & $(v=1,j=0)$ & 9.0 $\times$ 10$^{-14}$ & \cite{bodo03-jpc} \\
             & $(v=2,j=0)$ & 3.6 $\times$ 10$^{-12}$ & \cite{bodo03-jpc} \\
$^4$He + LiH & $(v=1,j=0)$ & 3.8 $\times$ 10$^{-13}$ & \cite{bodo03-jpc} \\
             & $(v=2,j=0)$ & 1.5 $\times$ 10$^{-11}$ & \cite{bodo03-jpc} \\ [1ex]
\hline
\end{tabular}
\end{center}
\caption{Zero-temperature inelastic rate coefficients for different atom - molecule systems.
\label{TAB1}
}
\end{table}

\subsubsection{Shape resonances in molecular collisions}

\begin{figure}[h]
\begin{center}
\includegraphics*[width=8cm,keepaspectratio=true,angle=0]{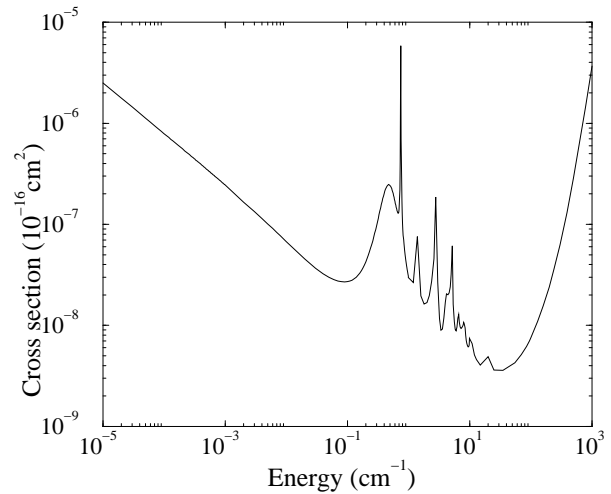}
\caption{Cross section for the quenching of the $v=1,j=0$ level of CO 
in collisions with $^4$He as a function of the incident kinetic energy.
Reproduced with permission  from Balakrishnan et al.~\cite{bala00}.
\label{HeCO-relaxation}
}
\end{center}
\end{figure}

At energies above the onset of the s-wave regime, cross sections will be dominated
by contributions from non-zero angular momentum partial
waves. If the interaction potential includes an attractive part, the 
effective potentials for non-zero angular momentum partial waves 
may possess centrifugal barriers that introduce shape resonances in the 
collision energy dependence of the cross section. This is illustrated 
in Fig.\ref{HeCO-relaxation} for the 
vibrational relaxation of CO($v=1,j=0$) in collisions with $^4$He atoms.

\begin{figure}[h]
\begin{center}
\includegraphics*[width=8cm,keepaspectratio=true,angle=0]{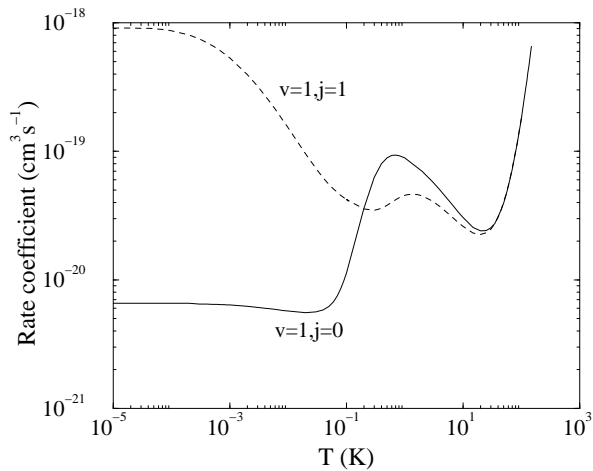}
\caption{Rate coefficients for the quenching of CO($v=1,j=0,1$) by collisions with $^4$He as functions of the 
temperature.
Reproduced  with permission from Balakrishnan et al.~\cite{bala00}.
\label{He-CO-rateconstant}
}
\end{center}
\end{figure}

The sharp features in the energy dependence of the cross section for energies between 
0.1 and 10.0 cm$^{-1}$ arises from shape resonances supported by the van der Waals 
interaction potential between He and the CO molecule. As shown in
 Fig.\ref{He-CO-rateconstant},
when integrated over the velocity distribution of the colliding species
 the shape resonances lead to significant enhancement of the vibrational
relaxation rate coefficient for temperatures between 0.1 and 10.0 K. 
Similar results have been found for vibrational relaxation of CO~\cite{zhu01}, 
O$_2$\cite{bala01b}, and CaH~\cite{bala03b} in collisions 
with $^3$He atoms. The sharp features in the energy dependence of the vibrational
relaxation cross sections for the Ar + D$_2$ sytem 
shown in Fig.\ref{Ar-D2-cross} also arises from shape resonances supported by the 
Ar$-$D$_2$ van der Waals potential.
The effect is generally more pronounced for systems composed of heavier diatomic
molecules and  
interaction potentials with deep van der Waals wells
for which the density of states will be much higher
leading to rich resonance structures in the cross sections.

\subsubsection{Feshbach resonances in molecular collisions}

\begin{figure}[h]
\begin{center}
\includegraphics*[width=8cm,keepaspectratio=true,angle=0]{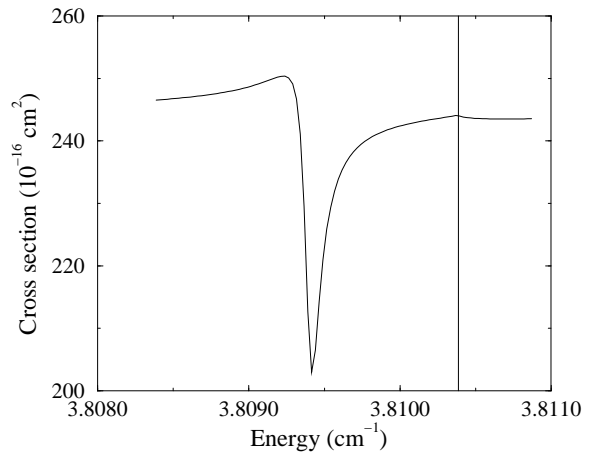}
\caption{Feshbach resonance in the elastic scattering cross section of CO($v=1,j=0$) by  
$^4$He atoms. The resonance occurs just below the opening of the $v=1,j=1$ level shown by the
vertical line. The energy is relative to the $v=1,j=0$ level of the CO molecule. Reproduced  with permission from Balakrishnan et al.~\cite{bala00}.
\label{HeCO-Feshbach}
}
\end{center}
\end{figure}

Feshbach resonances occur in multichannel scattering in which an unbound (continuum) channel
is coupled to a bound state of another channel. If the energy of the interacting system
in the  unbound channel lies close to that of the bound state
and the coupling between the two channels is strong the cross section may
change dramatically in the vicinity of the resonance. In the 
Feshbach resonance method for producing ultracold molecules, an external 
magnetic field is used to tune the energy of 
the bound pair
to that of the
the separated atoms.
In atom - diatom systems, the 
bound state may correspond to a quasibound state of the atom - diatom van der Waals
complex. Channel potentials corresponding to different initial vibrational and rotational
levels of the diatom may induce Feshbach resonances. For the He$-$CO system 
Feshbach resonances were found to occur near channel thresholds corresponding to the $j=1$ 
rotational level in the $v=0$ and $v=1$ vibrational levels. Fig.\ref{HeCO-Feshbach} shows
the Feshbach resonance in the elastic scattering cross sections in the $v=1,j=0$ channel
in the vicinity of the $v=1,j=1$ level. The presence of the Feshbach resonance 
close to the opening of the $j=1$ level has a dramatic effect
on the vibrational quenching cross sections from the $v=1,j=1$ level of the CO molecule.
Since the Feshbach resonance occurs so close to the threshold of the $v=1,j=1$ channel, 
its effect on scattering in the $v=1,j=1$ level is similar to that of the zero-energy
resonance discussed previously for the
Ar + D$_2$ system.
This is 
illustrated in Fig.~\ref{He-CO-rateconstant} (see also Table \ref{TAB1}) 
where we compare the rate coefficients for vibrational relaxation from the $v=1,j=0$
and $v=1,j=1$ levels of the CO molecule.
The zero-temperature limiting value of the quenching rate coefficient 
of the $v=1,j=1$ level is about two orders of magnitude larger than for the $v=1,j=0$ level.
Similar Feshbach resonances have also been shown to occur in the
vibrational and rotational predissociation of He$-$H$_2$ van der Waals complexes \cite{forrey98}. 
Forrey et al.\cite{forrey98} has successfully used the effective range theory to predict the predissociation 
lifetimes of these resonances.

The Feshbach resonances can be used as a very sensitive probe for the interaction
potential and also to selectively break or make bonds in chemical reactions.
The coupling between the bound and unbound states can be modified by 
applying an external electric or magnetic field and this provides an important
mechanism for creating or eliminating Feshbach resonances and thereby controlling the collisional
outcome. Krems have shown that weakly bound van der Waals complexes can be dissociated by tuning a 
Feshbach resonance using an external  magnetic field \cite{Krems04}. In this case the dissociation occurs through
coupling between Zeeman levels of the bound and unbound channels and the magnitude of the
coupling is varied by changing the external magnetic field.

\subsection{ Quasi-resonant transitions}

\begin{figure}[h]
\begin{center}
\includegraphics*[width=8cm,keepaspectratio=true,angle=0]{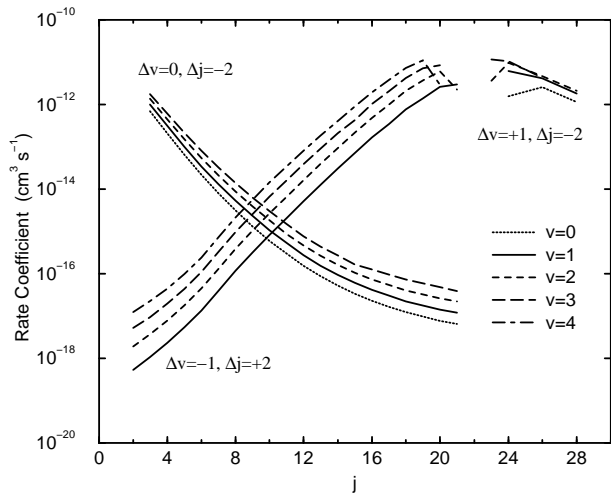} 
\caption{Zero-temperature rate coefficients for $^4$He + H$_2(v,j)$ collisions 
as functions of the initial vibrational and rotational quantum numbers. 
Reproduced with 
permission from Forrey et al.~\cite{forrey2001}.
\label{QR-transfer}
}
\end{center}
\end{figure}

While the properties of cold and ultracold collisions are quite different from scattering
 at thermal energies 
and quantum effects dominate at low temperatures, a remarkable correlation between classical and 
quantum dynamics has been discovered in the relaxation of ro-vibrationally excited diatomic molecules.
 Experiments performed nearly two decades ago~\cite{stewart88,magill88} showed that collisions of rotationally excited diatomic 
molecules with atoms may result in very efficient
 internal energy  transfer  between specific rotational and 
vibrational degrees of freedom. 
The energy transfer becomes highly efficient when the collision time is 
longer than the rotational period of the molecule. This effect has since been termed
``quasi-resonant rotation-vibration 
energy transfer". The experimental results revealed that the quasi-resonant (QR) transitions satisfy the
propensity rule $\Delta j=-4\Delta v$ or $\Delta j=-2\Delta v$ where $\Delta v=v_f-v_i$ and $\Delta j=j_f-j_i$~\cite{stewart88,magill88}. This 
inelastic channel dominates over all other ro-vibrational transitions. The QR transfer is 
generally insensitive to details of the interaction potential. Rather, the QR process involves 
conservation of the action, $I=n_vv+n_jj$, where $n_v$ and $n_j$ are small integers. Forrey et al.~\cite{forrey99a} found
 that the QR transitions also occur in cold and ultracold collisions of rotationally excited diatomic molecules with 
atoms and that the process is largely insensitive to the details of the interaction potential 
even in the ultracold regime. The $\Delta j=-2\Delta v$ QR transition in He + H$_2$ collisions~\cite{forrey2001} is 
illustrated in Fig.~\ref{QR-transfer} where the zero-temperature vibrational and 
rotational transition rate coefficients 
for different initial  vibrational levels of the H$_2$ molecule are 
plotted as functions of the initial rotational level. 
For initial rotational levels greater than 12 the QR transition becomes
the dominant energy transfer mechanism compared to pure rotational quenching. 
The gap at $j=22$ occurs because the 
$\Delta j=-2\Delta v$
 transition is energetically not accessible for this initial state at zero temperature. 
Forrey et al.~\cite{forrey99a}
 found that the QR process is even more dominant at low temperatures than at thermal energies. 
Remarkably, classical trajectory calculations~\cite{forrey99a}  were successful in correctly predicting the correlation
 between $\Delta j$ and $\Delta v$ at ultracold temperatures even though the changes in $v$ and $j$ were
 fractional. Extensive studies of QR energy transfer in cold and ultracold temperatures have been reported
 by Forrey et al.~\cite{forrey02,flasher02,florian04,mack06}. 
Ruiz and Heller~\cite{Heller06} have recently published a review paper providing 
a detailed analysis of QR phenomenon using semi-classical techniques. McCaffery and coworkers~\cite{maccaffery00,maccaffery02,marsh03} have also 
reported a number of quasi-classical trajectory calculations of the QR process 
in thermal energy collisions and successfully interpreted a
 large body of experimental data based on the QR phenomenon and simple parametric models.

\subsection{ Atom - molecular ion collisions}

Dynamics of ionic systems are different
from collisions of neutral species. 
The short-range part of the interaction potential for ionic systems
is usually more anisotropic and the long-range part
has an  attractive component which is determined by the polarizability  of the atom and it 
vanishes as $1/R^4$ where $R$ is the atom - molecule center-of-mass distance \cite{Bodo02}. Due to the 
strong polarizability term, the interaction potential for ionic systems extends to longer range compared 
to neutral atom - molecule systems. Therefore, it is important to understand the effect of
both the short and long range part of the interaction potential 
on the scattering dynamics.
For this purpose, several studies have focused on ultracold collisions between
molecular ions
and neutral atoms as well as neutral molecules
and atomic ions.

Bodo et al.~\cite{Bodo02}
investigated rotational quenching
in Ne$^+_2$ + Ne and He$^+_2$ + He collisions
at ultra-low energies.
They found that the Wigner regime 
begins at a collision energy of 10$^{-4}$~cm$^{-1}$ for 
the He system and 10$^{-6}$~cm$^{-1}$ for the Ne system.
In general, the s-wave Wigner regime was found to occur at lower energies for
ionic systems compared to
neutral species. 
For example, in He + H$_2$~\cite{bala98} and He + O$_2$~\cite{bala01b}
 collisions the Wigner regime begins at  collision energies of
about 10$^{-2}$~cm$^{-1}$.
The differences are attributed to  the long range of the ion - neutral interaction 
potential which enhances contributions from higher partial waves.
The differences between the He and Ne systems can be attributed to the mass 
difference and to the strength of the long range interaction potentials.
For the heavier Ne system  the long range part is more attractive, 
which increases the contribution of higher-order partial waves at ultra-low energies.
The  magnitude of the zero-energy rate coefficients for rotational quenching in these molecular ions
is on the order of 
10$^{-9}$~cm$^{3}$~s$^{-1}$, which is considerably larger than for collisions involving  
neutral species. \\

\begin{figure}[h]
\begin{center}
\begin{tabular}{cc}
\includegraphics*[bb=40 0 560 792,height=9cm,keepaspectratio=true,angle=-90]{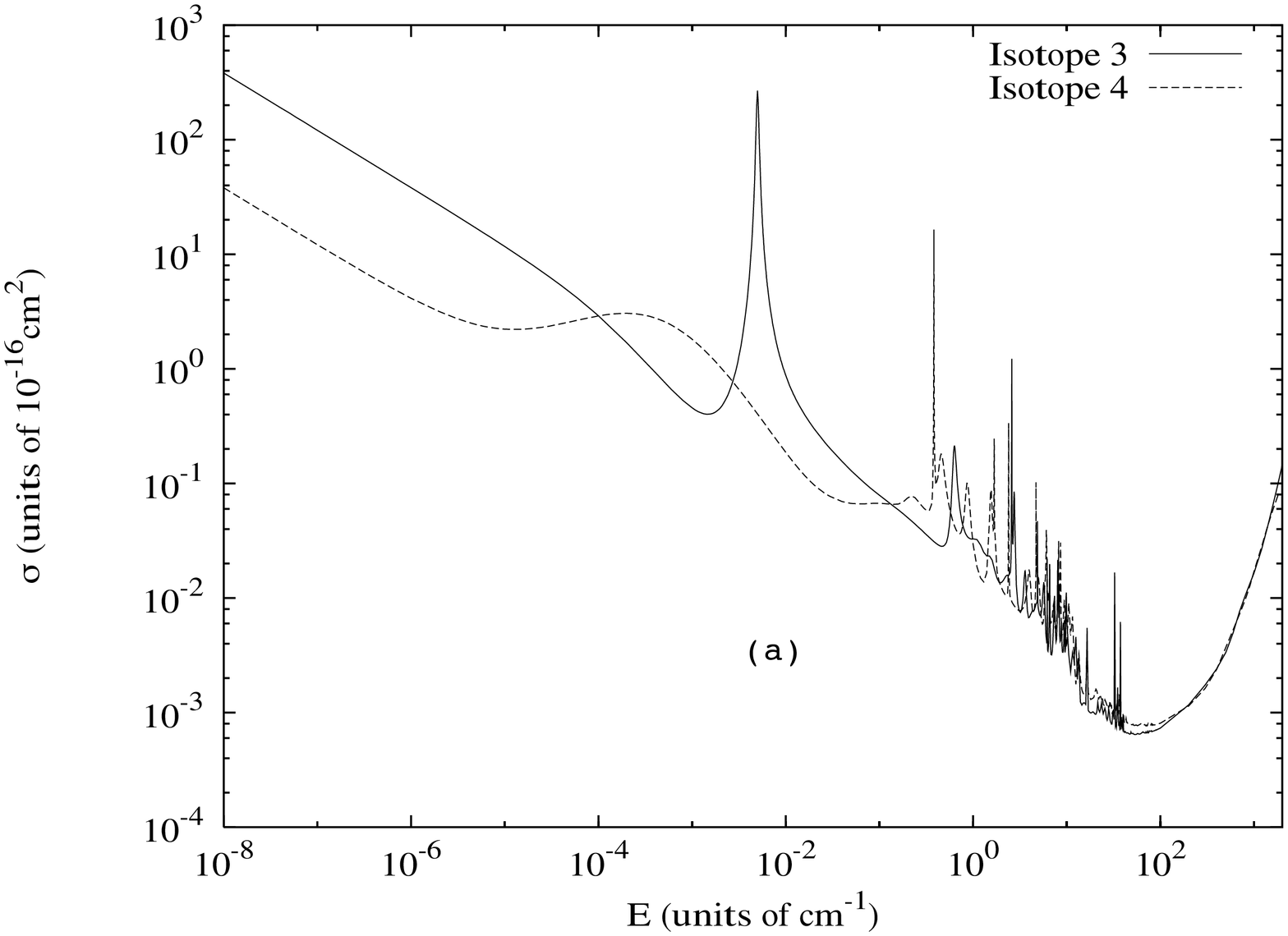} \\
\includegraphics*[bb=40 0 560 792,height=9cm,keepaspectratio=true,angle=-90]{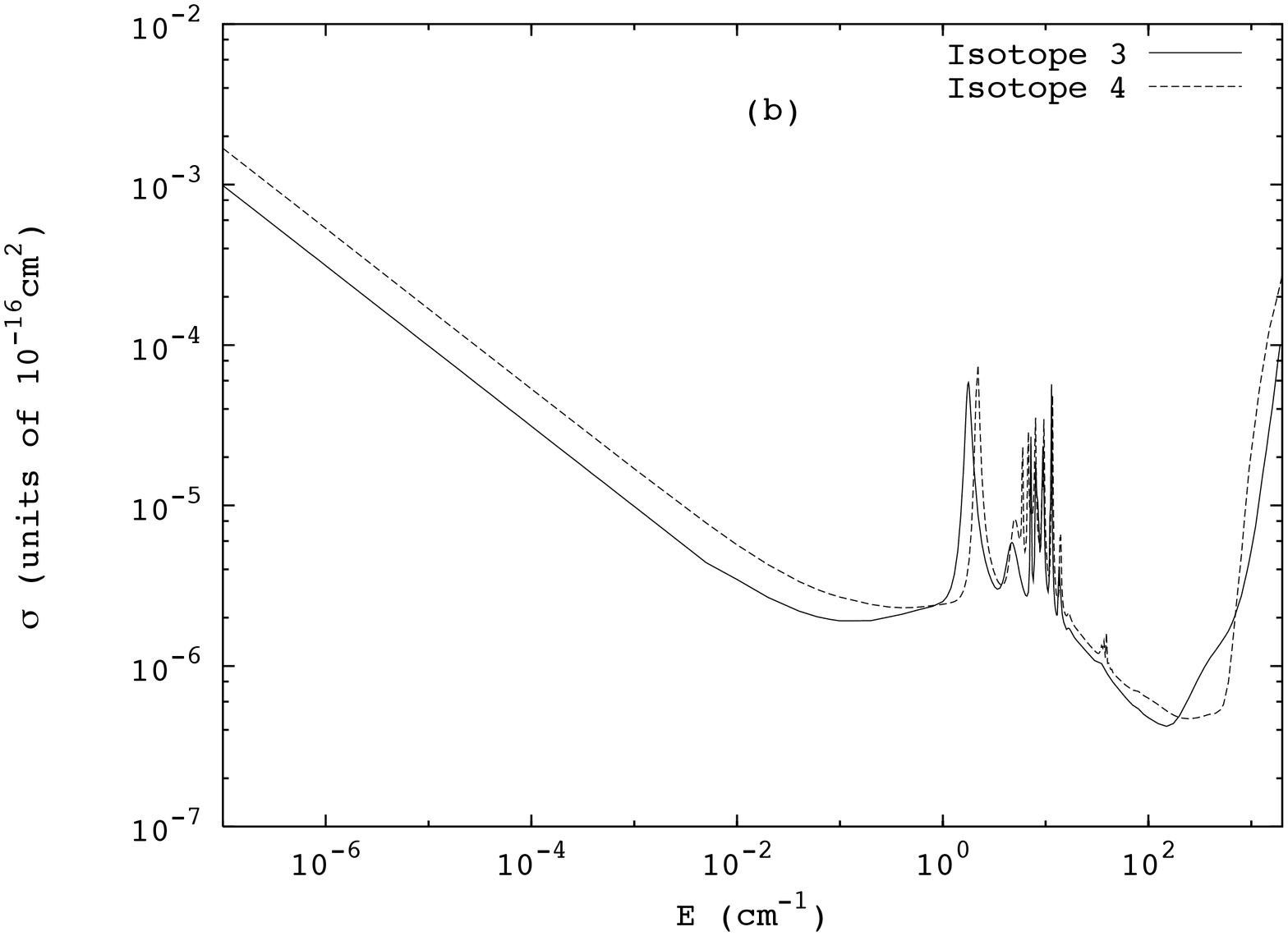} \\
\end{tabular}
\caption{Vibrational quenching cross sections for
N$^+_2(v=1,j=0)$ + $^{3}$He and N$^+_2(v=1,j=0)$ + $^{4}$He (upper panel)
and N$_2(v=1,j=0)$ + $^{3}$He and N$_2(v=1,j=0)$ + $^{4}$He (lower panel) systems.
Reproduced with permission from Stoecklin et al.~\cite{Stoecklin05}.
\label{N2HE-FIG}
}
\end{center}
\end{figure}

Stoecklin et al.~\cite{Stoecklin05}
have performed a comparative 
study of collisions of N$^+_2(v=1,j=0)$ molecular ions 
 with neutral $^{3}$He or $^{4}$He atoms,
and the scattering of neutral N$_2(v=1,j=0)$ molecules
in collisions with $^{3}$He or $^{4}$He atoms.
The vibrational quenching cross sections for these systems
are presented in Fig.~\ref{N2HE-FIG}. 
While the behavior of the quenching cross sections
of neutral N$_2$ molecules with $^{3}$He and $^{4}$He atoms was found to be similar,
they observed a striking difference between the quenching cross sections 
of N$^+_2$ in collisions with $^{3}$He and $^{4}$He atoms.
In particular, the resonance positions in the quenching cross sections were 
significantly shifted and  the number of resonances was different.
The N$^+_2$ + $^{3}$He system has a shape resonance at 
a collision energy of 10$^{-2}$~cm$^{-1}$
which is absent for the N$^+_2$ + $^{4}$He system.  
Furthermore, 
the zero-energy quenching rate coefficient
is an order of magnitude larger
for collisions of N$^+_2(v=1,j=0)$ with $^{3}$He
than with $^{4}$He,
whereas it is comparable
for collisions of N$_2(v=1,j=0)$ with both $^{3}$He
and $^{4}$He.
The differences
may be due 
to a virtual state in the N$^+_2(v=1,j=0)$ + $^{3}$He
collision system.
More recently, Guillon et al.~\cite{Guillon07}
studied the effect of spin-rotation interaction
on vibrational and rotational quenching for the He$-$N$^+_2$ system.
They found that  the vibrational quenching is not modified
by the spin-rotation coupling, while
rotational transitions are sensitive to the fine-structure interactions. \\

Table~\ref{TAB2} provides a compilation of zero-temperature quenching rate coefficients for 
vibrational and rotational relaxation in several ion atom - molecule systems. 

\begin{table}[h]
\begin{center}
\begin{tabular}{c c c c}
\hline
system & initial $(v,j)$ & $k_{T=0}$ (cm$^3$s$^{-1}$) & Ref.  \\ [0.5ex]
\hline
Ne$_2^+$ + Ne & $(v=0,j=2)$ & 3.3 $\times$ 10$^{-10}$ & \cite{Bodo02} \\
              & $(v=0,j=4)$ & 7.3 $\times$ 10$^{-10}$ & \cite{Bodo02} \\
              & $(v=0,j=6)$ & 7.0 $\times$ 10$^{-10}$ & \cite{Bodo02} \\
\hline
He$_2^+$ + He & $(v=0,j=2)$ & 6.7 $\times$ 10$^{-10}$ & \cite{Bodo02} \\
              & $(v=0,j=4)$ & 8.4 $\times$ 10$^{-10}$ & \cite{Bodo02} \\
              & $(v=0,j=6)$ & 1.2 $\times$ 10$^{-9}$ & \cite{Bodo02} \\
\hline
N$_2^+$ + $^3$He & $(v=1,j=0)$ & 4 $\times$ 10$^{-14}$ & \cite{Stoecklin05} \\
N$_2^+$ + $^4$He & $(v=1,j=0)$ & 3 $\times$ 10$^{-15}$ & \cite{Stoecklin05} \\ [1ex]
\hline
\end{tabular}
\end{center}
\caption{Zero-temperature inelastic rate coefficients for different ion atom - molecule systems.
\label{TAB2}
}
\end{table}

\section{Chemical Reactions at ultracold temperatures}

The last three decades have seen impressive progress in the experimental and theoretical descriptions of 
chemical reactions between atoms and small molecular systems. While fully state-resolved experiments
have been performed for a large number of collision systems, accurate quantum dynamics calculations have
been restricted to systems involving light atoms such as H + H$_2$, F + H$_2$, Cl + H$_2$, C + H$_2$, N + H$_2$,
O + H$_2$ and some of their 
isotopic analogues ~\cite{Althorpe03,Hu06}. Most experimental studies of small molecular
 systems focussed on reactions at thermal 
or elevated collision energies though some recent measurements have been extended to 
temperatures as low as 10~K for some astrophysically relevant systems~\cite{Smith03}. 
Due to the possibility of achieving coherent chemistry
there has been substantial interest in understanding the behavior of chemical
reactions at cold and ultracold temperatures. 
At these temperatures perturbations introduced by external electric
and magnetic fields are significant compared to the collision energies involved and external
fields can be employed to control and manipulate the reaction outcome.
Over the last seven years a number of studies of ultracold atom - diatom
chemical reactions have been reported both for reactions with and without an energy barrier.
Chemical reactions at low temperatures often behave quite differently from reactions at elevated
temperatures. In particular, 
the weak van der Waals interaction potential which does not play any
significant role at high temperatures may have
a dramatic effect on the outcome of reactions at low
temperatures.

\subsection{ Tunneling dominated reactions}

In the course of the last
$10-15$ years there has been much interest in understanding 
the role of resonances in chemical
reactions that involve an energy barrier for which the reactivity is primarily driven by tunneling at low temperatures. 
Recent studies on F + H$_2$/HD/D$_2$~\cite{bala-cpl-2001,castillo96,takayanagi98,skodje00,bodo2002,bala03a,bodo04,qui06,aldegunde06,lee06,tao07,fazio07},
Cl + HD~\cite{skouteris99,balakrishnan04c}, H + HCl/DCl~\cite{weck04},
Li + HF / LiF + H~\cite{weck05a,weck05b}, O + H$_2$~\cite{weck05c,weck06a}, and 
F + HCl/DCl~\cite{goulven08} reactions
have demonstrated that decay of quasibound states of the van
der Waals interaction potential in the entrance and 
exit channels may give rise to narrow Feshbach resonances in the
cross sections.

The reactions of F with H$_2$ and HD have been the subject of numerous quantum scattering calculations over the 
last two decades. The two reactions have emerged as benchmark systems for experimental and
 theoretical studies of resonances in chemical reactions. A detailed analysis of the low energy resonance
 features in the F + H$_2$ reaction was presented by Castillo et al. in 1996~\cite{castillo96}. They 
demonstrated that several resonances that appear in the energy dependence of the cumulative reaction probability for the
F + H$_2$ reaction arise due to the van der Waals interaction potential in the product HF + H
channel. In particular, the origin of the resonances has been attributed to the van der Waals potential 
associated with the HF($v'=3, j'=0-3$)  channels. A large number of 
experimental and theoretical papers has appeared which examined various aspects of these resonance
features~\cite{bala-cpl-2001,castillo96,takayanagi98,skodje00,bodo04,qui06,aldegunde06,lee06,tao07,fazio07}.
  Among these studies, the work of Takayanagi and Kurosaki~\cite{takayanagi98} deserves special attention. They 
showed that for F + H$_2$, F + HD and F + D$_2$ reactions reactive scattering resonances occur due to the decay 
of quasibound states of the van der Waals potential in the entrance channel of the reaction. These
 Feshbach resonances are associated with the decay of quasibound states of adiabatic potentials 
corresponding to F$\cdots$H$_2(v=0,j=0,1)$, 
F$\cdots$HD$(v=0,j=0-2)$, and F$\cdots$D$_2(v=0,j=0-2)$ complexes, obtained
 by diagonalizing the $J=0$ Hamiltonian in a basis set of the asymptotic ro-vibrational states of the reactant molecules.

\subsubsection{ Reactions at zero temperature}

\begin{figure}[h]
\begin{center}
\includegraphics*[width=8cm,keepaspectratio=true,angle=0]{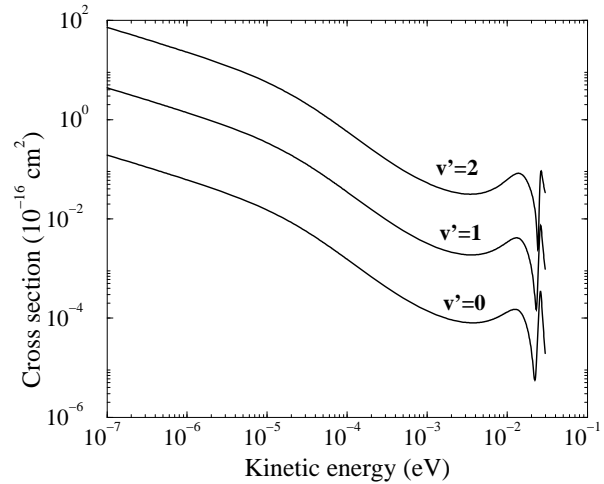}
\caption{$J=0$ cross sections for  F + H$_2(v=0,j=0)$ $\to$ HF($v'$) + H reaction for $v'=0-2$ as functions of
the incident kinetic energy.
Reproduced  with permission from Balakrishnan and Dalgarno~\cite{bala-cpl-2001}.
\label{F-H2-Cross}
}
\end{center}
\end{figure}

Balakrishnan and Dalgarno~\cite{bala-cpl-2001} showed that quasibound states of the F + H$_2$ van der Waals
 complex have a dramatic effect on the reactivity in the Wigner threshold regime. They found that the 
F + H$_2(v=0,j=0)$ reaction has a rate coefficient of $1.25 \times 10^{-12}$~cm$^3$~s$^{-1}$ 
in the zero-temperature limit. 
Their calculation was based on the widely used PES of Stark and Werner~\cite{stark96} for the
F + H$_2$ system. The relatively
large value of the zero-temperature rate coefficient is due to the presence of a 
small narrow energy barrier for the reaction so that tunneling of the H atom is efficient. The vibrational level
resolved cross sections 
for the F + H$_2(v=0,j=0)$ reaction 
are shown in Fig.~\ref{F-H2-Cross} for the total angular momentum
quantum number $J=0$. The $v'=2$ level is the 
dominant channel at low energies in agreement with the behavior at higher energies. \\

\begin{figure}[h]
\begin{center}
\includegraphics*[width=8cm,keepaspectratio=true,angle=0]{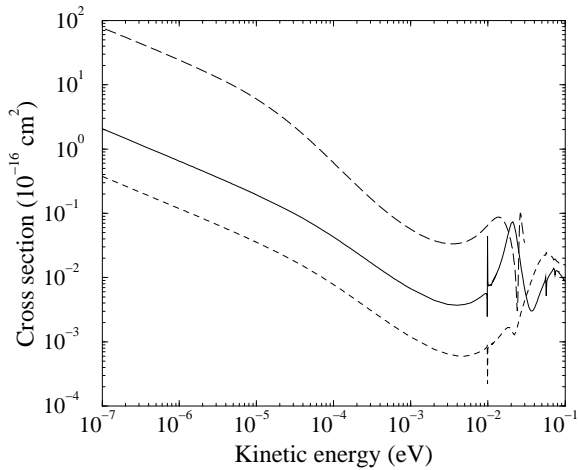} 
\caption{Comparison of $J=0$ cross sections for F + HD($v=0,j=0$) $\to$ HF + D (solid curve), 
F + HD($v=0,j=0$) $\to$ DF + H (short-dashed curve) 
and F + H$_2(v=0,j=0)$ $\to$ HF + H (long-dashed curve) reactions
as functions of the incident kinetic energy.
Reproduced  with permission from Balakrishnan and Dalgarno~\cite{bala03a}.
\label{FHD-Cross}
}
\end{center}
\end{figure}

Subsequent calculations showed that at low energies the F + HD 
reaction is dominated by the formation
of the HF product with an HF/DF branching ratio of about 5.5~\cite{bala03a}. The formation of the DF product is
suppressed because tunneling of the heavier D atom is less efficient.
In earlier quantum calculations, Baer and coworkers~\cite{baer99a,zhang00} 
 reported HF/DF branching ratios
ranging from 1.5 at 450~K to 6.0 at 100~K. Their low temperature value is in close agreement with the zero-temperature 
limit obtained by Balakrishnan and Dalgarno~\cite{bala03a}.
Fig.~\ref{FHD-Cross} compares the $J=0$
cross sections for the F + H$_2(v=0,j=0)$ and F + HD($v=0,j=0$) reactions over a kinetic energy range of $10^{-7}$ to 1.0~eV. 
In the Wigner threshold regime,
the reactivity of the F + H$_2$ system is an order of magnitude greater than that of the F + HD reaction. 
A rigorous analysis of the scattering resonances in the F + HD reaction was recently presented by De Fazio
et al.~\cite{fazio07}, who provided a detailed characterization of the resonances supported by the entrance and exit 
channels of the van der Waals potential and discussed the effect of higher total angular momenta on the position 
and lifetime of the resonances.
Stereodynamical aspects of the F + H$_2$ collision and the effect of polarization of the H$_2$
 molecule on the outcome of the  reaction at low energies were recently explored by Aldegunde et al.~\cite{aldegunde06}. They argued 
that a reactant polarization scheme can be exploited to control state-to-state dynamics of the reaction. \\

\begin{figure}[h]
\begin{center}
\begin{tabular}{cc}
\includegraphics*[height=11cm,keepaspectratio=true]{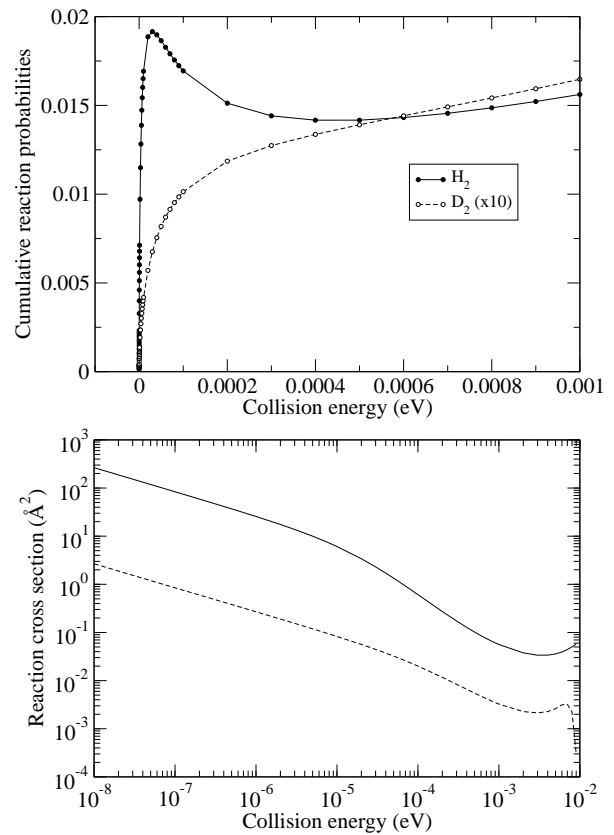} 
\end{tabular}
\caption{Comparison of the $J=0$ cumulative reaction probabilities (upper panel) and  
cross sections (lower sub-panel) of F + H$_2$ and F + D$_2$ reactions as functions of the
incident kinetic energy.
Reproduced  with permission from Bodo et al.~\cite{bodo04}.
\label{F-H2-D2}
}
\end{center}
\end{figure}

To explore the role of tunneling in chemical reactions at cold and ultracold temperatures
Bodo, Gianturco, and Dalgarno~\cite{bodo2002} investigated the dynamics of the F + D$_2$ system at low and ultra-low
energies. They found that compared to the F + H$_2$ reaction, the reactivity of 
F + D$_2$  is significantly suppressed in the Wigner regime with
an HF/DF ratio of about 100. This is illustrated in Fig.~\ref{F-H2-D2} where  
the $J=0$ cumulative reaction probability and 
cross sections for the F + H$_2$ and F + D$_2$ reactions are plotted as functions of the incident kinetic energy
from the 
Wigner limit to 0.01~eV.

The dramatic suppression of the F + D$_2$ reaction in the 
Wigner regime cannot be explained based on tunneling alone. A closer examination of the reaction probabilities for the F + H$_2$ and F + D$_2$ 
reaction revealed that an unusual enhancement of the reactivity occurs for the F + H$_2$ reaction at about
$3 \times 10^{-5}$~eV~\cite{bodo04} (see the upper panel of Fig.~\ref{F-H2-D2}). 
This feature was attributed to the presence of a virtual state. 
The enhancement of the 
limiting value of the rate coefficient for the F + H$_2$ reaction was also ascribed to the virtual state. Evidence
for the virtual state is the presence of a Ramsauer--Townsend minimum in the elastic cross 
section at an energy of about $3 \times 10^{-5}$~eV and a negative value of the real part of 
the scattering length for the F + H$_2$ reaction \cite{bodo04}. 

\begin{figure}[h]
\begin{center}
\begin{tabular}{cc}
\includegraphics*[height=11cm,keepaspectratio=true]{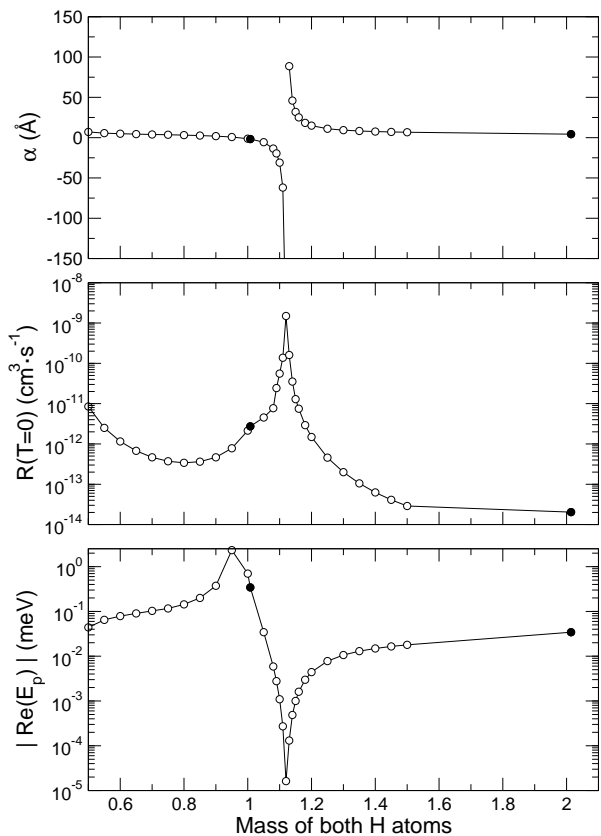}
\end{tabular}
\caption{Real part of the scattering length (upper panel), zero-temperature limiting values of the rate
coefficient (middle panel), and positions of quasibound states (lower panel) of the F$\cdots$H$_2$ complex
as functions of the mass of a pseudo hydrogen atom.
Reproduced  with permission from Bodo et al.~\cite{bodo04}.
\label{F-H2-D2-scaled}
}
\end{center}
\end{figure}

To explore how isotope substitution modifies reactivity at
 low temperatures Bodo et al.~\cite{bodo04} 
artificially varied the mass of the hydrogen atoms in the calculation for the F + H$_2$
reaction from 0.5 to 1.5~amu. As illustrated in Fig.~\ref{F-H2-D2-scaled}, for an H atom 
mass of 1.12~amu the virtual state
induces a zero energy resonance at which the real part of the scattering length diverged to
infinity. For the same value of the H atom mass, the zero-temperature rate coefficient 
of the reaction attains a 
value of $1.0\times 10^{-9}$~cm$^3$~s$^{-1}$ which is about three orders of magnitude larger than that of the F + H$_2$
reaction in the Wigner limit. The variation of the scattering length of the F + H$_2$ system as a function
of the mass of the pseudo-hydrogen atom is similar to the variation of scattering length as a function of
the magnetic field in the vicinity of a Feshbach resonance.\\

\begin{figure}[h]
\begin{center}
\includegraphics*[width=8cm,keepaspectratio=true,angle=0]{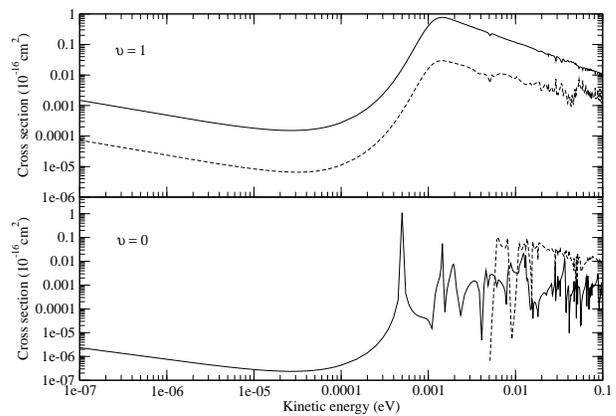} 
\caption{Cross sections for LiF formation (solid curve) and non-reactive scattering (dashed curve) in Li + HF($v,j=0$) 
collisions as functions
of the incident kinetic energy: results for $v=0$ (lower panel); results for $v=1$ (upper panel).  
Reproduced  with permission from Weck and Balakrishnan~\cite{weck05a}.
\label{Li-HF-Cross}
}
\end{center}
\end{figure}

Another example of a tunneling dominated reaction is 
the Li + HF $\to$ LiF + H reaction. At cold and ultracold temperatures the reaction 
occurs by tunneling of 
the relatively heavy fluorine atom. The LiH + F channel is energetically not accessible at low energies and tunneling of the H atom
is not involved in this reaction at low temperatures. Due to strong
electric dipole forces exerted by the HF molecule, the van der Waals interaction potential of the LiHF
system is deeper compared to the F + H$_2$ system. The Li$\cdots$HF van der Waals potential
well is about 0.24~eV (1936.0~cm$^{-1}$) and the H$\cdots$LiF potential well is about 0.07~eV (565.0~cm$^{-1}$).
Quantum scattering calculations for Li + HF and LiF + H reactions by Weck and Balakrishnan~\cite{weck05a,weck05b} 
with the PES of Aguado et al.~\cite{agua03} have shown that for 
incident energies below 10$^{-3}$~eV the reaction cross section exhibits a large number of resonances.
The energy dependence of the $J=0$ cross section for Li + HF($v=0,j=0$) collisions is shown in Fig.~\ref{Li-HF-Cross}.
 Detailed bound state calculations of the LiHF van der Waals complexes revealed that for the Li + HF($v=0,j=0$)
 reaction the resonances correspond to the decay of  
Li$\cdots$HF($v=0,j=1-4$) van der Waals complexes. Calculations with vibrationally excited HF molecules showed that the reaction 
becomes about 600 times more efficient in the Wigner regime when HF is excited to the $v=1$ vibrational level.
 As seen in Fig.~\ref{Li-HF-Cross}, a unique feature of 
the Li + HF($v=0,j=0$) reaction is the presence of a strong peak at 
$5.0 \times 10^{-4}$~eV at which the reaction cross section is about six orders of magnitude larger than the 
background cross section.  The results for both Li + HF~\cite{weck05a} and LiF + H~\cite{weck05b} reactions with thermal and non-thermal 
vibrational excitation suggest that heavy-atom tunneling may play an important role in chemical reactions
 at cold and ultracold temperatures. 
An experimental study~\cite{zuev03} of an organic ring expansion
reaction at 8~K has shown that the reaction occurs almost exclusively by carbon tunneling.
The tunneling contribution was found to be orders of magnitude
 greater than over the barrier contribution.
In a recent work, Tscherbul and Krems \cite{tscherbul08} have explored the Li + HF and LiF + H reactions
in the presence of an external electric field. They have shown that, for temperatures below 1 K,
the reaction probability can be significantly influenced by electric fields.

\subsubsection{ Feshbach resonances in reactive scattering}

Van der Waals complexes formed during collisions can either undergo vibrational predissociation
or vibrational prereaction leading to sharp features in the energy dependence of the cross
sections. The term
``prereaction" refers to the process in which a rotationally or vibrationally excited van der Waals 
complex decays through chemical reaction rather than by rotational or vibrational predissociation. For reactions with energy barriers the chemical reaction pathway
may involve tunneling.
The reactions of Cl with H$_2$ and HD  are dominated by tunneling at low 
temperatures~\cite{skouteris99,balakrishnan04c,weck04}. Compared 
to the F + H$_2$ and F + HD reactions, the energy barrier for the Cl + H$_2$ reaction 
is much larger and the reactivity at low energies is
significantly suppressed. Nearly a decade ago, Skouteris et al.~\cite{skouteris99} showed that 
the van der Waals interaction potential between Cl and HD determines the reaction outcome, 
despite the fact that 
the depth of the
 van der Waals interaction potential for the Cl$\cdots$HD system is less than one-tenth of the height of 
the reaction barrier. 
Quantum scattering calculations of the Cl + HD reaction on PESs
without the van der Waals interaction potential predict nearly equal probabilities for HCl and DCl 
products~\cite{skouteris99}. However, if the potential surface includes the van der Waals interaction a strong preference for 
the DCl product occurs at thermal energies, in agreement with experimental results. 
The effect of these 
weakly bound states on the reactivity at cold and ultracold temperatures has recently been explored by
Balakrishnan~\cite{balakrishnan04c}.

\begin{figure}[h]
\begin{center}
\includegraphics*[width=8cm,keepaspectratio=true,angle=0]{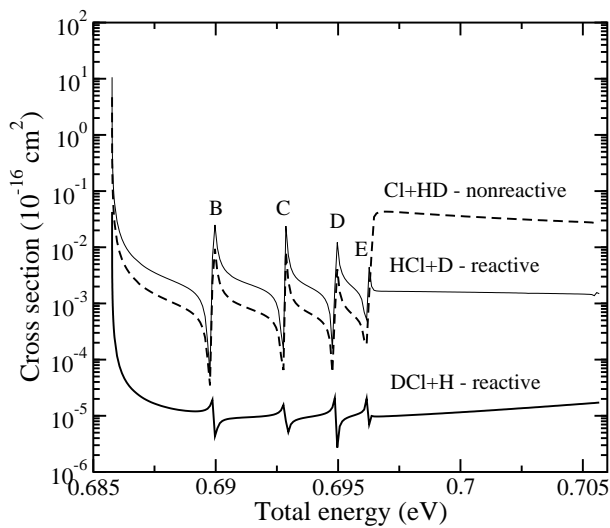} 
\caption{Reactive and non-reactive scattering cross sections for Cl + HD($v=1,j=0$) reaction as functions of the 
incident kinetic energy.
Reproduced  with permission from Balakrishnan~\cite{balakrishnan04c}.
\label{Cl-HD-resonances}
}
\end{center}
\end{figure}

\begin{figure}[h]
\begin{center}
\includegraphics*[width=8cm,keepaspectratio=true,angle=0]{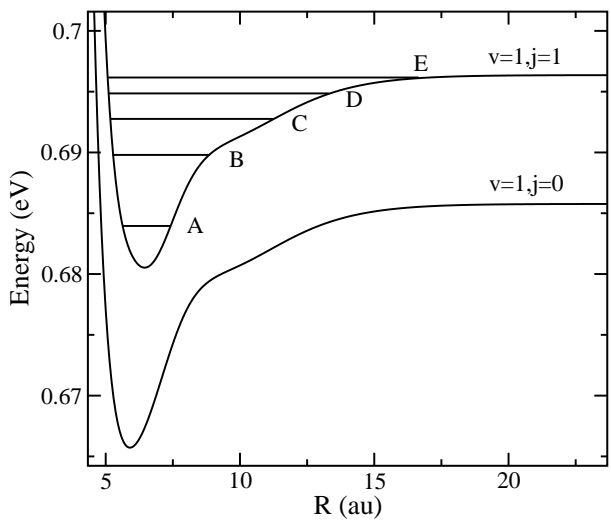} 
\caption{Adiabatic potential energy curves of the Cl + HD system correlating with the 
HD($v=1,j=0$) and HD($v=1,j=1$) levels as functions of the atom - molecule separation.
Quasibound levels responsible for the resonances in Fig.\ref{Cl-HD-resonances} are labeled 
by B, C, D, and E.
Reproduced  with permission from Balakrishnan~\cite{balakrishnan04c}.
\label{Cl-HD-adiabatic-potentials}
}
\end{center}
\end{figure}

The cross sections for HCl and DCl formation and non-reactive rovibrational transitions
in Cl + HD($v=1,j=0$) collisions for total angular momentum quantum number $J=0$ are shown in
Fig.~\ref{Cl-HD-resonances} as functions of the total energy~\cite{balakrishnan04c}. 
The sharp
features in the cross sections correspond to Feshbach resonances arising from the
decay of quasibound van der Waals complexes formed in the entrance channel of the reaction.
The quasibound states can be identified by examining bound states of adiabatic 
potentials correlating with the $v=1,j=0$ and $v=1,j=1$ levels of the HD molecule. They
are displayed in Fig.~\ref{Cl-HD-adiabatic-potentials} as functions of the atom - molecule
separation. The adiabatic potential curves are computed by diagonalizing the diabatic potential
energy matrix obtained in a basis set of rovibrational levels of the HD molecule at each
value of the atom - molecule separation.
The Feshbach resonances labeled B, C, D, and E in Fig.~\ref{Cl-HD-resonances} result from the 
decay of 
the corresponding quasibound states shown in Fig.~\ref{Cl-HD-adiabatic-potentials}.

\begin{figure}[h]
\begin{center}
\includegraphics*[width=8cm,keepaspectratio=true,angle=0]{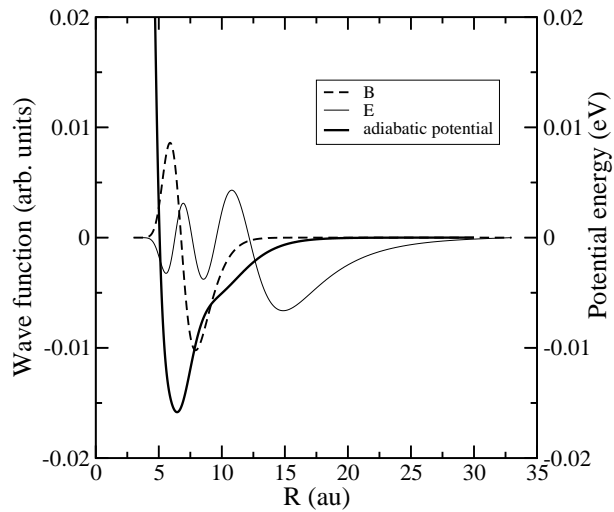} 
\caption{Adiabatic potential and the 
wavefunctions of the quasibound levels B and E shown in 
Fig.\ref{Cl-HD-adiabatic-potentials} as functions of the atom - molecule separation. 
Amplitudes of the wavefunctions have been reduced by a factor of 10 for the convenience 
of plotting.
Reproduced  with permission from Balakrishnan~\cite{balakrishnan04c}.
\label{Cl-HD-wavefunctions}
}
\end{center}
\end{figure}

The cross sections in Fig.~\ref{Cl-HD-resonances} do not show a
 peak corresponding to the 
metastable state A. This is because the state A is too deeply bound and it
 is not accessible through scattering in the $v=1,j=0$ channel. Fig.~\ref{Cl-HD-resonances}
also shows that the quasibound states preferentially undergo prereaction than
predissociation. This is due to the enhanced coupling of 
the resonance states with the reactive channels 
compared to those with the non-reactive channels. The wavefunctions of quasibound states B and E are 
shown in Fig.~\ref{Cl-HD-wavefunctions} as functions of the atom - molecule separation. 
Although the wavefunction of the weakly bound state E extends far beyond the transition
state region of the reaction, it preferentially
undergoes prereaction compared to predissociation. Thus, regions of the interaction
potential far away from the transition state region may have a significant effect on
reactivity especially when part of the wavefunction in that region is 
sampled by a resonance state which is coupled to the reactive channel. Fig.~\ref{Cl-HD-resonances}
also demonstrates that
since the 
reaction is dominated by tunneling at low temperatures the formation of the DCl + H product which involves
 the tunneling of the D atom is severely suppressed in the threshold regime. This is
more clearly illustrated in Fig.~4 of Ref.~\cite{balakrishnan04c} where  the reaction cross section 
is plotted as a function of the 
incident kinetic energy. \\

Table~\ref{TAB3} provides a compilation of zero-temperature quenching rate coefficients for 
a number of atom - diatom chemical reactions which are dominated by tunneling at low energies.

\begin{table}[h]
\begin{center}
\begin{tabular}{c c c c}
\hline
system & initial $(v,j)$ & $k_{T=0}$ (cm$^3$s$^{-1}$) & Ref.  \\ [0.5ex]
\hline
F + H$_2$ & $(v=0,j=0)$ & 1.3 $\times$ 10$^{-12}$ (H + HF) & \cite{bala-cpl-2001} \\
F + HD & $(v=0,j=0)$ & 2.8 $\times$ 10$^{-14}$ (D + HF)  & \cite{bala03a} \\
       & $(v=0,j=0)$ & 0.5 $\times$ 10$^{-14}$ (H + DF)  & \cite{bala03a} \\
F + D$_2$ & $(v=0,j=0)$ & 2.1 $\times$ 10$^{-14}$ (D + DF)& \cite{bodo02-cpl} \\
\hline
F + HCl & $(v=0,j=0)$ & 1.2$\times$ 10$^{-17}$ (Cl + HF) & \cite{goulven08} \\
F + HCl & $(v=1,j=0)$ & 4.0$\times$ 10$^{-15}$ (Cl + HF) & \cite{goulven08} \\
        &             & 5.3$\times$ 10$^{-15}$ (F + HCl) & \cite{goulven08} \\
F + HCl & $(v=2,j=0)$ & 5.2$\times$ 10$^{-13}$ (Cl + HF) & \cite{goulven08} \\
        &             & 3.3$\times$ 10$^{-13}$ (F + HCl) & \cite{goulven08} \\
F + DCl & $(v=1,j=0)$ & 4.4$\times$ 10$^{-21}$ (Cl + DF) & \cite{goulven08} \\
\hline
H + HCl & $(v=0,j=0)$ & 2.4$\times 10^{-19}$ (Cl + H$_2$) & \cite{weck04}\\
        & $(v=1,j=0)$ & 7.2$\times 10^{-14}$ (Cl + H$_2$) & \cite{weck04}\\
        & $(v=2,j=0)$ & 1.9$\times 10^{-11}$ (Cl + H$_2$) & \cite{weck04}\\
H + DCl & $(v=1,j=0)$ & 7.8$\times 10^{-15}$ (Cl + HD) & \cite{weck04}\\
        & $(v=2,j=0)$ & 1.7$\times 10^{-12}$ (Cl + HD) & \cite{weck04}\\
\hline
Cl + HD & $(v=1,j=0)$ & 1.7 $\times$ 10$^{-13}$ (D + HCl) & \cite{balakrishnan04c} \\
        &             & 7.1 $\times$ 10$^{-16}$ (H + DCl) & \cite{balakrishnan04c} \\
        &             & 7.8 $\times$ 10$^{-14}$ (Cl + HD) & \cite{balakrishnan04c} \\
\hline
Li + HF & $(v=0,j=0)$ & 4.5 $\times$ 10$^{-20}$ (H + LiF) & \cite{weck05a} \\
        & $(v=1,j=0)$ & 2.8 $\times$ 10$^{-17}$ (H + LiF) & \cite{weck05a} \\
H + LiF & $(v=1,j=0)$ & 3.8 $\times$ 10$^{-15}$ (Li + HF) & \cite{weck05b} \\
        &             & 1.6 $\times$ 10$^{-14}$ (H + LiF) & \cite{weck05b} \\
        & $(v=2,j=0)$ & 1.7 $\times$ 10$^{-14}$ (Li + HF) & \cite{weck05b} \\
        &             & 2.7 $\times$ 10$^{-13}$ (H + LiF) & \cite{weck05b} \\ [1ex]
\hline
\end{tabular}
\end{center}
\caption{Zero-temperature quenching rate coefficients for tunneling dominated reactions.
The final arrangement has also been specified.
\label{TAB3}
}
\end{table}

\subsection{ Barrierless reactions}

\subsubsection{Collision systems of three alkali metal atoms}

As discussed by Hutson and Sold\'{a}n in 
two recent reviews~\cite{Hutson06,Hutson07a}, 
progress 
on  the production of ultracold molecules
and the creation of molecular Bose--Einstein condensates of alkali-metal systems
have motivated theoretical studies on ultracold
atom - dimer alkali-metal collisions.
Here, we give an overview of recent calculations for spin-polarized triatomic alkali-metal systems,
Li + Li$_2$~\cite{Cvitas05a,Cvitas05b,Quemener07,Cvitas07}, 
Na + Na$_2$~\cite{Soldan02,Quemener04} and K + K$_2$~\cite{Quemener05}. 
These results have all been obtained using the reactive scattering code 
written by Launay and Le Dourneuf~\cite{Launay89}, based on a time-independent
quantum formalism.
Refs.~\cite{Cvitas05a,Cvitas05b,Cvitas07,Quemener05}
describe the details of the PESs and dynamics calculations.
The quantum dynamics studies of Refs.~\cite{Soldan02,Quemener04}
have been performed using the Na$_3$ PES reported by Higgins et al.~\cite{Higgins00},
while the results of Ref.~\cite{Quemener07} have been obtained using
the Li$_3$ PES calculated by Colavecchia et al.~\cite{Colavecchia03}.
Another PES for Li$_3$ has been constructed by Brue et al.~\cite{Brue05}.

Quantum scattering calculations show that ultracold reactions of alkali metal atoms 
with alkali metal dimers are much more efficient than tunneling driven processes. 
This can be explained based on the following considerations.
First, the indistinguishability of the atoms may play a role.
For a homonuclear triatomic system, 
the three possible arrangement channels are the same.
Also the presence of  identical two-body potentials in all three arrangement channels may
 enhance the
depth of the  three-body interaction  potential. 
Consider the two-body term (additive term) of a homonuclear triatomic system 
at equilateral geometries.
Since the distances between the three identical atoms are the same,
the same diatomic potential term is added three times to yield the two-body terms
of the triatomic system.
If the diatomic potential is deep or repulsive, 
the two-body term will be three times deeper or repulsive
at equilateral geometries.
In contrast, for triatomic systems with distinguishable atoms,
the diatomic pairs are not the same.
For example, for the Li + HF system,  
the three diatomic fragments are distinctly different.
They have completely different electronic structures and properties such as
minima, equilibrium distances, turning points, and the nature and range of the interaction.
Thus, at an equilateral configuration
for which the three diatomic distances are the same but the diatomic potential energies
are different (one may be attractive while the other two may be repulsive),
the overall two-body term can be weaker than the individual two-body interaction potential.
This could lead to a triatomic interaction region that is less strong than that of a homonuclear triatomic system.

Second, the topology of the PES plays a significant role.
For all alkali-metal trimer systems, the minimum energy configurations arise at
equilateral and  collinear geometries as shown by Sold{\'a}n et al.~\cite{Soldan03}. 
Furthermore, the surface is barrierless so that 
all atom - dimer alkali-metal collisional approaches are energetically possible. 
In contrast, most of the non-alkali systems discussed above are characterized by
a collinear (or bend) atom - diatom approach, and  dominated by a repulsive
barrier in the triatomic transition-state region. 
The particular topology of the PES arises
from the three-body interaction potential.
Also, the couplings between different electronic surfaces
of the triatomic system give rise to conical intersections, which
create a repulsive barrier in the transition-state region.
If the energy barrier is high and its width large, tunneling will be very
inefficient leading to very small values for the reaction rate coefficients in the Wigner regime.  
Although reactivity can be enhanced when resonances are present, 
the background scattering in reactive cross sections is generally quite small.
In contrast, for almost all alkali-metal trimer systems, the conical intersections arise
at small interatomic distances where the PES is sufficiently repulsive
that it plays no significant role at ultra-low energy and the
reactivity is not influenced by such repulsive barriers.

Third, the density of states of the system plays an important role.
If the density of states is small as for light systems, 
the typical energy spacing is rather large.
For heavy systems such as alkali-metal systems, 
the density of states is very large and the narrow 
energy spacing can lead to very strong couplings,
especially in the vicinity of avoided crossings.
The strong couplings will lead to efficient energy transfer between different
quantum states.
This may also explain the relatively weak dependence of the vibrational quenching rate 
coefficients on the initial vibrational state of the molecule in atom - dimer alkali-metal collisions. \\

\begin{figure}[h]
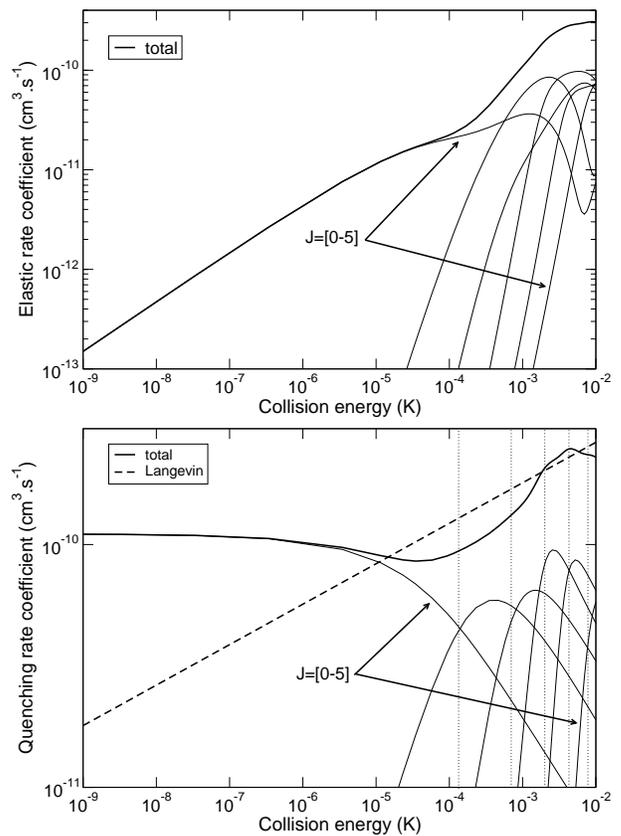

\begin{center}
\begin{tabular}{cc}
\includegraphics*[width=8cm,keepaspectratio=true]{Figure17a.eps} \\ 
\includegraphics*[width=8cm,keepaspectratio=true]{Figure17b.eps}
\end{tabular}
\caption{Elastic (upper panel) and quenching (lower panel) 
rate coefficients versus the collision energy
for $^{39}$K + $^{39}$K$_2(v=1,j=0)$ scattering.
The rate coefficient
from a capture model
is also shown for the quenching processes.
Adapted  with permission from Qu{\'e}m{\'e}ner et al.~\cite{Quemener05}.
\label{RAELINK3-FIG}
}
\end{center}
\end{figure}

Fig.~\ref{RAELINK3-FIG} provides a comparison between
 elastic and quenching rate coefficients
for $^{39}$K + $^{39}$K$_2(v=1,j=0)$~\cite{Quemener05}
collisions at energies ranging from
10$^{-9}$~K to 10$^{-2}$~K.
The results include contributions from the total angular momentum quantum numbers $J=0-5$.
For low vibrational states of the K$_2$ dimer,
quenching processes are more efficient
than elastic scattering at ultracold energies. 
For example, at a temperature of $10^{-9}$~K, 
the quenching rate coefficient is about $10^{-10}$~cm$^3$~s$^{-1}$ 
compared to $10^{-13}$~cm$^3$~s$^{-1}$ for  elastic scattering.
The quenching processes lead to collisional relaxation of the molecules
to lower ro-vibrational states resulting in trap loss.

Similar results have been obtained for 
Na + Na$_2$~\cite{Soldan02,Quemener04} and Li + Li$_2$~\cite{Cvitas05a,Quemener07,Cvitas07} collisions.
The zero-energy quenching rate coefficients for these systems are found to be on the order of 
$10^{-11}-10^{-10}$~cm$^3$~s$^{-1}$. 
The quenching processes are more efficient 
than the elastic collisions 
at ultracold temperatures.
Thus, quenching of vibrationally excited  alkali-metal dimers in collisions with alkali-metal atoms
occurs with significant rates  at ultracold temperatures.
The typical  magnitude of these rate coefficients from the scattering calculations 
is in reasonable  agreement
with experimental results.
In a recent experiment, Staanum et al.~\cite{Staanum06}
reported a value of  $9.8 \times 10^{-11}$~cm$^3$~s$^{-1}$
for the relaxation of low-lying vibrational levels of 
Cs$_2$ in collisions with Cs atoms at a temperature of $60 \times 10^{-6}$~K.
In a separate experiment, Zahzam et al.~\cite{Zahzam06}
determined a quenching rate coefficient of
$2.6 \times 10^{-11}$~cm$^3$~s$^{-1}$ for the same system
at a temperature of $40 \times 10^{-6}$~K.
Wynar et al.~\cite{Wynar00}
estimated inelastic rate coefficients of $8 \times 10^{-11}$~cm$^3$~s$^{-1}$
for Rb + Rb$_2$ collisions.
Mukaiyama et al.~\cite{Mukaiyama04}
reported an inelastic rate coefficient of $5.5 \times 10^{-11}$~cm$^3$~s$^{-1}$
for collisions of Na atoms with Na$_2$ molecules created by the Feshbach resonance method while
Syassen et al.~\cite{Syassen06} obtained a value
of $2 \times 10^{-10}$~cm$^3$~s$^{-1}$ for collisions of 
Rb atoms with Rb$_2$ molecules.\\

At large atom - diatom separations, $R$, the interaction potential can be approximated by 
an effective potential, composed of a
repulsive centrifugal term and the long range interaction potential.
For alkali-metal trimer systems, the atom - diatom long range potential 
is a van der Waals interaction potential
and behaves as ~-~C$_6/R^6$. 
The classical capture model (also known as the Langevin model) has been shown 
to work quite well for these systems \cite{Quemener05} at certain energy regimes. 
The Langevin model 
is  based on the assumption 
that if the total energy of the system
can overcome the effective barrier, then the elastic probability is zero
and the quenching probability is one.
Otherwise, the elastic probability is one and the quenching probability is zero.
Classically, the barrier prevents the atom and the molecule 
from  accessing the region of strong coupling with the other channels.

Results obtained using the Langevin model for the K + K$_2$ system are shown in 
Fig.~\ref{RAELINK3-FIG} along with the quantum results.
As Fig.~\ref{RAELINK3-FIG} illustrates,
three different regions can be distinguished for the K + K$_2$ system.
The first region, for collision energies below 10$^{-6}$~K, corresponds to the Wigner regime
where the threshold laws apply. In this regime, the quenching rate coefficient 
tends to a constant while the elastic component approaches zero as the square root of the collision energy.
A quantum description is required 
and classical models cannot describe the dynamics in this regime.
The second region, above a collision energy of 10$^{-3}$~K, corresponds to the
Langevin regime.  
Here, the difference between the full quantum calculation and the classical model
is within 10~\%.
Similar results have been found for $^7$Li + $^7$Li$_2(v=1,2,j=0)$ 
and  $^6$Li + $^6$Li$_2(v=1,2,j=1)$ collisions by Cvita{\v s} et al.~\cite{Cvitas05a},
who showed that in the Langevin regime the rate coefficients become independent
of the vibrational level of the molecule. 
The third region is
intermediate  between the Wigner and Langevin regimes, which typically corresponds
to a quantum calculation with  two or three partial waves.
As shown in  the lower panel of Fig.~\ref{RAELINK3-FIG}  the height of each 
$J$-resolved effective barrier depicted by vertical lines for $J=1-5$
corresponds approximately to the maximum
of each $J$-resolved quenching rate coefficient.
This indicates that the quenching process
becomes significant when the barrier height is energetically overcome.
Whenever the elastic and quenching rate coefficients have comparable values,
 the Langevin regime is reached, as shown in Fig.~\ref{RAELINK3-FIG}.
Since the quenching probability is almost equal to one
and the elastic probability is almost equal to zero,
elastic and quenching transition matrix elements are both equal to one
and the cross sections and rate coefficients for the two processes become similar.

\begin{figure}[h]
\begin{tabular}{c}
\includegraphics*[width=8cm,keepaspectratio=true]{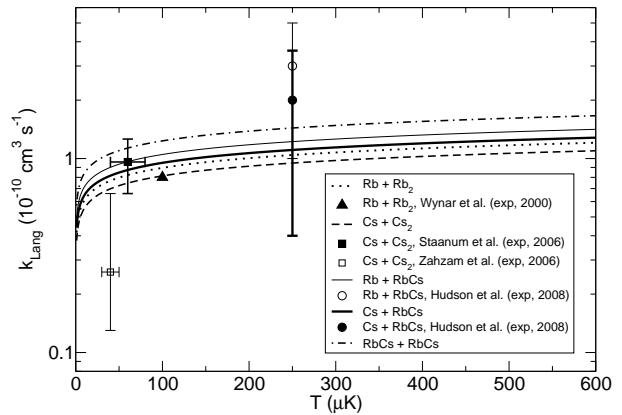}
\end{tabular}
\caption{Rate coefficients as  functions of the temperature
predicted by the Langevin model compared with experimental data
 for different collisional processes
involving Rb or Cs atoms. The curves correspond to the Langevin results and the symbols denote
the experimental results.
\label{LANGEVIN-RBCS}
}
\end{figure}

The above analysis shows that 
at energies above the onset of the Wigner regime the Langevin model can correctly describe the dynamics for 
barrierless atom - dimer alkali-metal collisions because of the strong inelastic couplings.
However, the classical model is not suitable for tunneling dominated reactions. 
Accurate quantum dynamics calculations are computationally demanding for heavier systems such as
Cs + Cs$_2$, Rb + Rb$_2$, Rb + RbCs or Cs + RbCs and the classical model may be used to
qualitatively describe the dynamics of these systems at energies above the s-wave regime.
The temperature dependence of 
the rate coefficients predicted by the Langevin model 
is given by the simple formula in atomic units:
\begin{eqnarray*}
k_{\text{Lang}}(T)= \pi \left( \frac{8 k_B T}{\pi \mu} \right)^{1/2}
\left( \frac{2 C_6}{k_B T} \right)^{1/3} \Gamma(2/3)
\end{eqnarray*}
where $\mu$ is the reduced mass of the  atom - diatom collisional system
and $C_6$ the dominant atom - diatom long range coefficient.
In Fig.~\ref{LANGEVIN-RBCS} we show the rate coefficients predicted by the Langevin model
as  functions of the temperature
for different atom - diatom and diatom - diatom collisions
involving Rb or Cs atoms.
For Rb$-$RbCs, Cs$-$RbCs, and RbCs$-$RbCs collisions we used the 
corresponding $C_6$
coefficients calculated by Hudson et al.~\cite{Hudson08}. In the absence of similar data for
Cs + Cs$_2$ and Rb + Rb$_2$ collisions, we approximated the atom - dimer
$C_6$ coefficient by multiplying the corresponding atom - atom value by a factor of two.
For Cs + Cs$_2$ collisions we used the $C_6$ coefficient for Cs$-$Cs interaction calculated by
Amiot et al.~\cite{Amiot02} and for Rb + Rb$_2$ collisions we adopted the $C_6$ coefficient for
Rb$-$Rb interaction
reported by Derevianko et al.~\cite{Derevianko99}.
As shown in Fig.~\ref{LANGEVIN-RBCS} the Langevin model predicts rate coefficients within an order of magnitude
of the experimental values for the different systems.
The predicted results for the 
Cs + Cs$_2$, Rb + RbCs, and Cs + RbCs collisions
agree with the corresponding experimental data of Refs.~\cite{Staanum06,Zahzam06,Hudson08}
within the reported  error bars.
For Rb + Rb$_2$ collisions the Langevin model predicts results in
 very close agreement with the experimental result of Wynar et al.~\cite{Wynar00}.
Thus, the Langevin model  appears to be valid for describing collisional properties of
 these systems at $\mu$K temperatures. 
The model also confirms that cold and ultracold collisions of alkali metal atoms and dimers
are essentially characterized by the leading term in the long-range part of the interaction potential.

\subsubsection{Role of PES in determining ultracold reactions}

Potential energy surfaces are the key ingredients that enter in quantum dynamics calculations.
While the two-body terms are generally well known
and accurate, 
the three-body terms are
 more difficult to compute with high precision
since they are non-additive and involve correlations between the three atoms.
As a consequence, quantum dynamics calculations may suffer from 
the quality and degree of accuracy of the three-body terms.
The sensitivity of the ultracold collision cross sections to the details of  the PES has been investigated
for atom - dimer alkali-metal systems 
by Qu\'{e}m\'{e}ner et al.~\cite{Quemener04}
for  Na + Na$_2(v=1-3,j=0)$
and by Cvita\v{s} et al.~\cite{Cvitas07}
for  Li + Li$_2(v=0-3,j=0)$.
In these studies, a linear scaling factor $\lambda$
has been included to tune  the three-body interaction term in the PES and
 cross sections were calculated as a function of $\lambda$.
Other studies have compared the dynamics
with  and without the three-body terms
for Na + Na$_2(v=1,j=0)$~\cite{Soldan02}
and for Li + Li$_2(v=0-10,j=0)$~\cite{Quemener07} collisions.

\begin{figure}[h]
\includegraphics*[width=8cm,keepaspectratio=true]{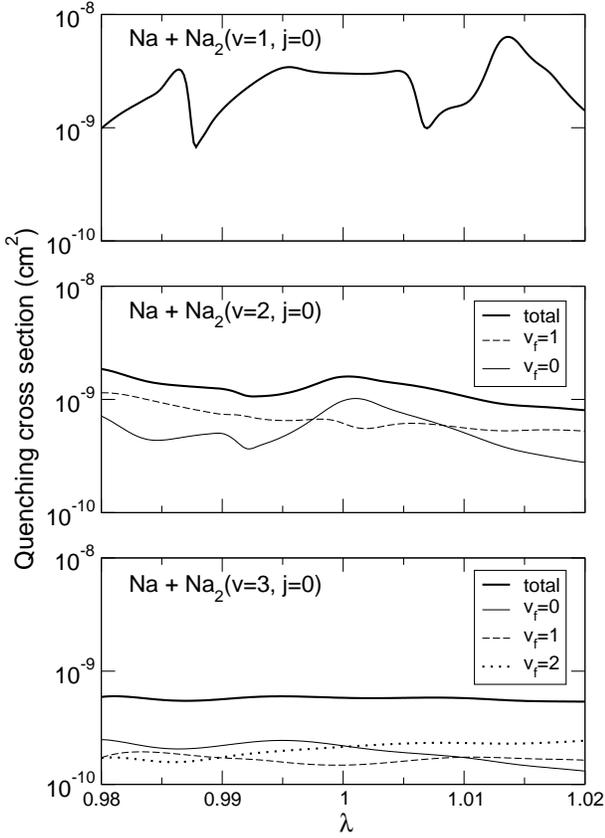}
\caption{Dependence of the quenching cross sections on the three-body term
at a collision energy of 10$^{-9}$~K 
for Na + Na$_2(v=1-3,j=0)$.
Reproduced  with permission from Qu\'em\'ener et al.~\cite{Quemener04}.
\label{SENSIB-FIG}
}
\end{figure}

The dependence of the total 
quenching cross sections
on the three body term for  
Na + Na$_2(v=1-3,j=0)$ scattering~\cite{Quemener04}
at a collision energy of 10$^{-9}$~K is shown in
Fig.~\ref{SENSIB-FIG}.
The contribution of each final vibrational level
is also plotted.
For the $v=1$ vibrational level, the cross sections are very sensitive
to the details of the three-body potential.
A change of 1~\% in $\lambda$ leads to a significant change
of 75~\% in the cross sections. At the minimum of the Na$_3$ potential,
a change of 1~\% corresponds approximately to 10~K.
At present, ab initio calculations of PESs cannot be done with such accuracy.
Thus it appears that it is difficult to get an accuracy better than two orders of magnitude
in the cross sections for
Na + Na$_2(v=1,j=0)$ collisions.
However, for $v=2$ and 3,
the total cross sections show a weaker dependence on the three-body term.
The 
state-resolved cross sections do not show  strong dependence on the three-body term,
compared to collisions of molecules in 
$v=1$.

\begin{figure}[h]
\includegraphics*[bb=40 20 550 380,width=8.0cm,keepaspectratio=true]{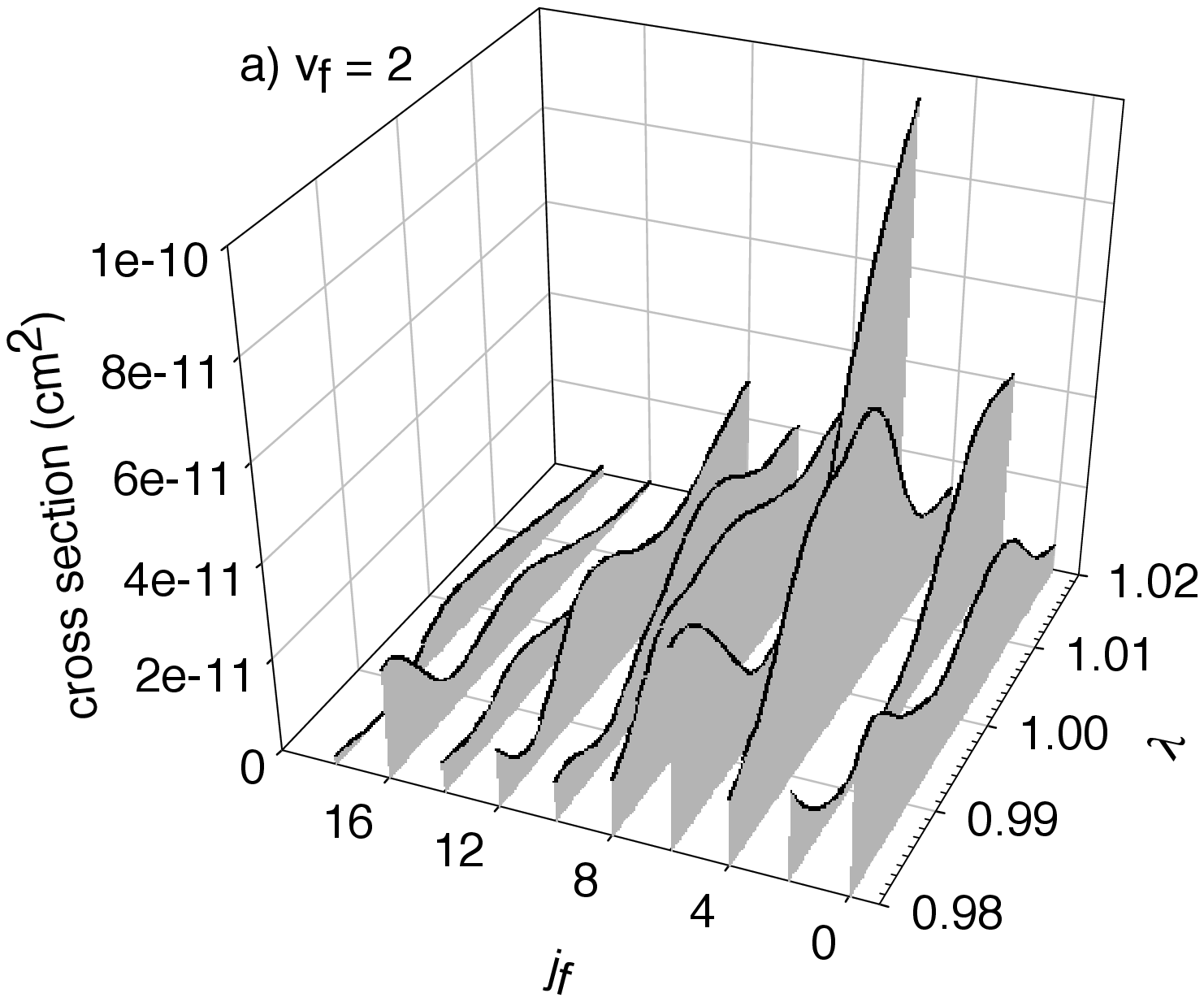} \\
\includegraphics*[bb=40 20 550 380,width=8.0cm,keepaspectratio=true]{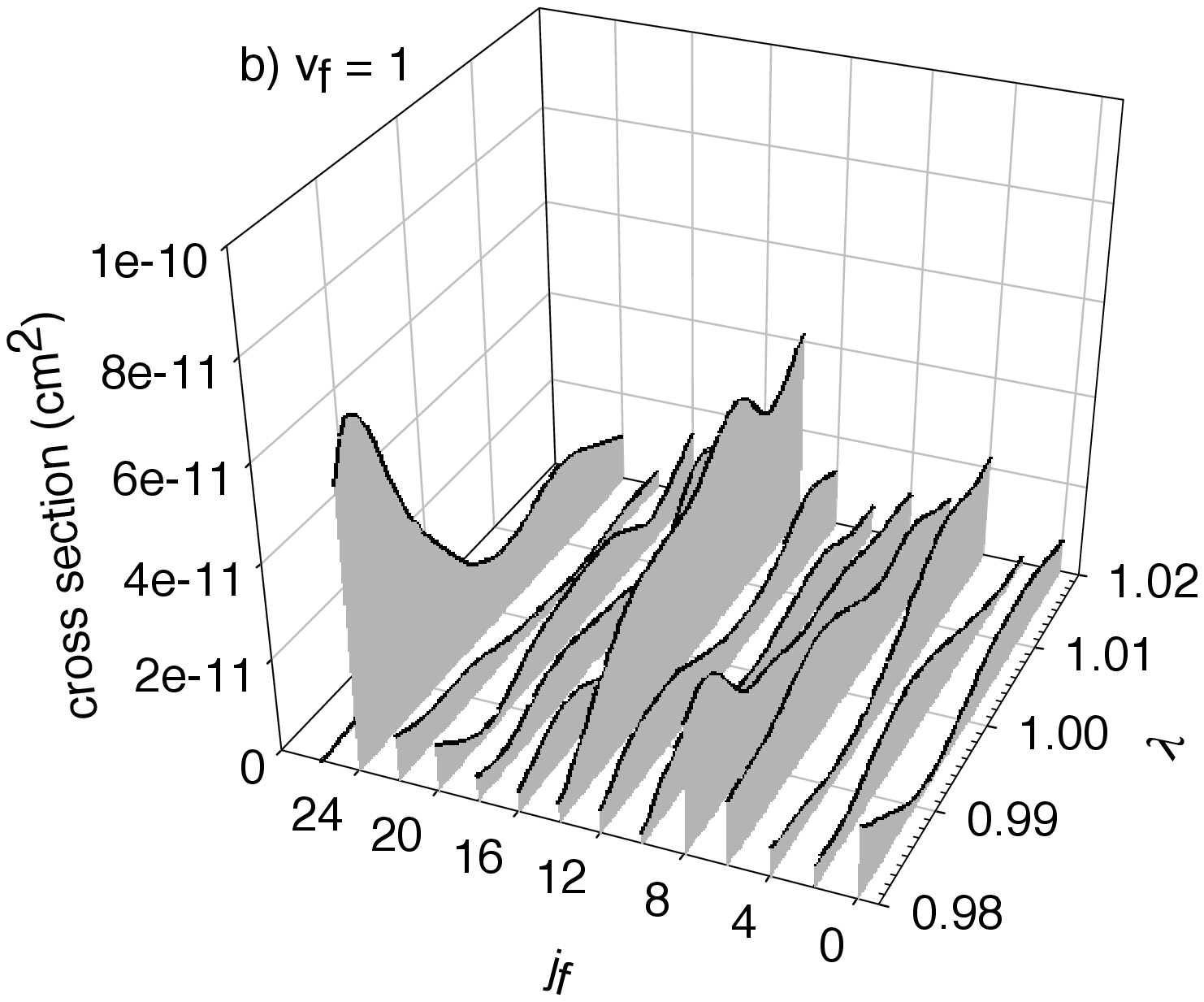} \\
\includegraphics*[bb=40 20 550 380,width=8.0cm,keepaspectratio=true]{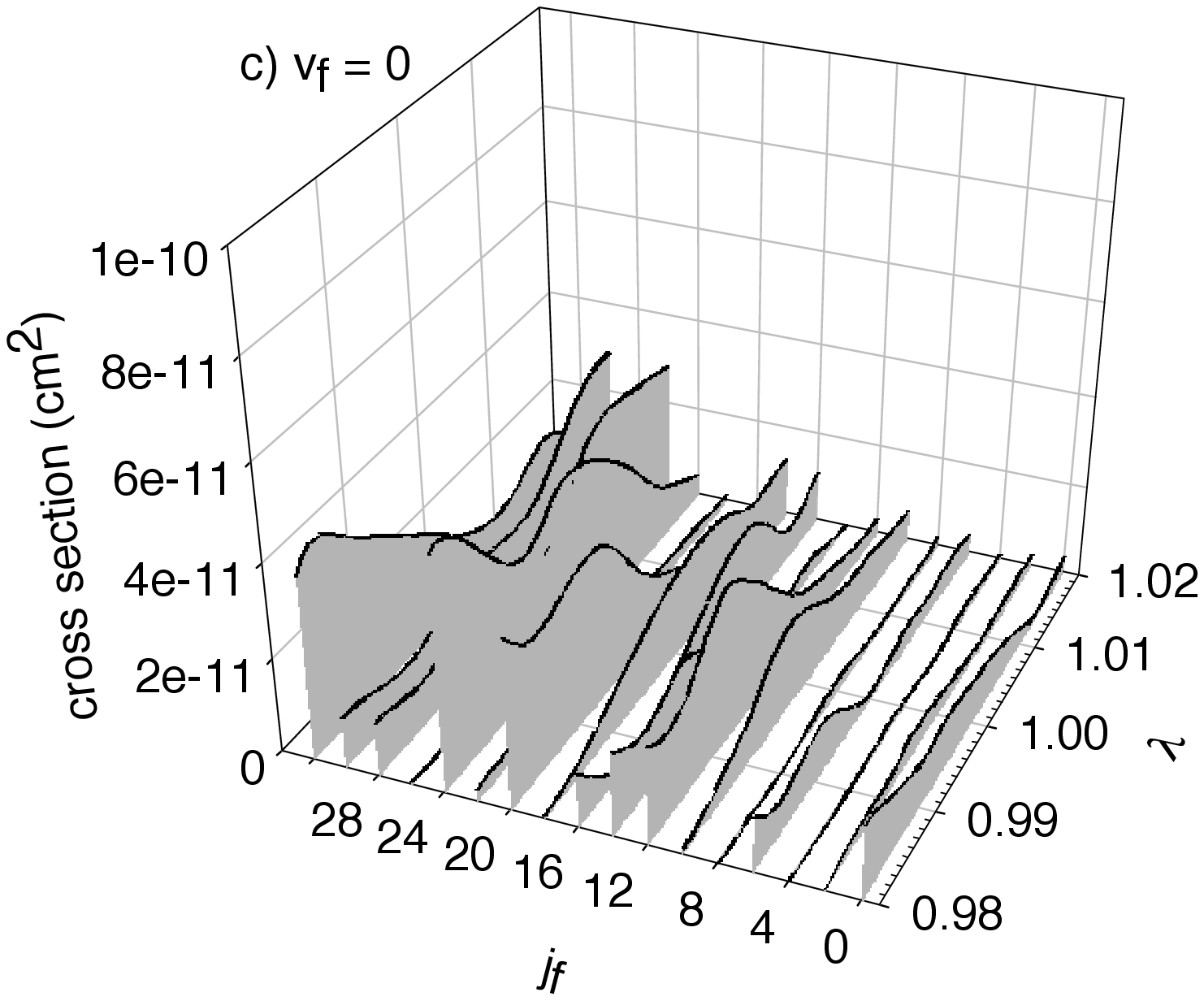}
\caption{Variation of state-to-state cross sections at a 
collision energy of 10$^{-9}$~K
for Na + Na$_2(v=3,j=0)$ $\to$ Na + Na$_2(v_f=2,1,0,j_f)$ 
as a function of the parameter $\lambda$. See text for details.
Reproduced  with permission from Qu\'em\'ener et al.~\cite{Quemener04}.
\label{SENSIB-FIG-2}
}
\end{figure}

Fig.~\ref{SENSIB-FIG-2} shows the dependence of the 
state-to-state 
cross sections
on the three body term for 
the reaction
Na + Na$_2(v=3,j=0)$  $\to$ Na + Na$_2(v_f=2,1,0,j_f)$ 
at a collision energy of 10$^{-9}$~K.
The oscillations in the cross sections, when $\lambda$ is modified,
are due to Feshbach resonances which
arise when a triatomic quasibound state (or a  virtual state) crosses the energy threshold 
as the strength of the interaction potential is decreased (increased).
When such a Feshbach resonance occurs,
Cvita\v{s} et al.~\cite{Cvitas07} have argued,
strong resonant peaks appear in the cross sections
if inelastic couplings are weak.
In contrast, weak oscillations of 
one order of magnitude at best appear 
when inelastic couplings
are strong, as in alkali-metal trimer systems. 
A generalization of this effect has been discussed by Hutson~\cite{Hutson07b}. 
Larger modifications of the three-body term affect the cross sections significantly.
This can also be seen in Fig.~\ref{RATEallv-LI3-FIG}
for Li + Li$_2$ collisions~\cite{Quemener07} in which the three-body term is excluded
from the dynamics calculation.
The state-to-state  cross sections
exhibit a stronger dependence on the three body term.

\subsubsection{Relaxation of vibrationally excited alkali metal dimers}

Ultracold  diatomic molecules produced in photoassociation or Feshbach resonance methods are
usually created in excited vibrational states.
Theoretical studies involving highly vibrationally excited molecules are challenging due to the 
large number of  energetically open reaction channels present in the quantum calculation. 
This puts severe restriction on the calculations for vibrationally excited molecules at low temperatures.
Ultracold quantum dynamics calculations for collisions of highly vibrationally excited molecules 
have been reported in 2007 by Qu\'em\'ener et al.~\cite{Quemener07} for the  
Li + Li$_2$ system. 
A three-atom problem is generally described by
 two different kinds of asymptotic channels.
The first kind are the single-continuum states (SCSs). 
They correspond to configurations where the triatomic system
dissociates asymptotically into an atom and
a diatomic molecule. 
The second kind are the double-continuum states (DCSs).
They correspond to 
configurations where the triatomic system
dissociates asymptotically into three separated  atoms.
Since highly  vibrationally excited molecular states 
lie close to and below the triatomic dissociation limit,
they are also coupled with the DCSs which lie above the dissociation limit.  
As a consequence, DCSs have to be included in quantum simulations of 
atom - diatom systems involving highly vibrationally excited diatomic molecules~\cite{Quemener07}.
This dramatically  increases  the size and complexity of the quantum dynamics problem. \\

\begin{figure}[h]
\begin{center}
\includegraphics*[width=8cm,keepaspectratio=true]{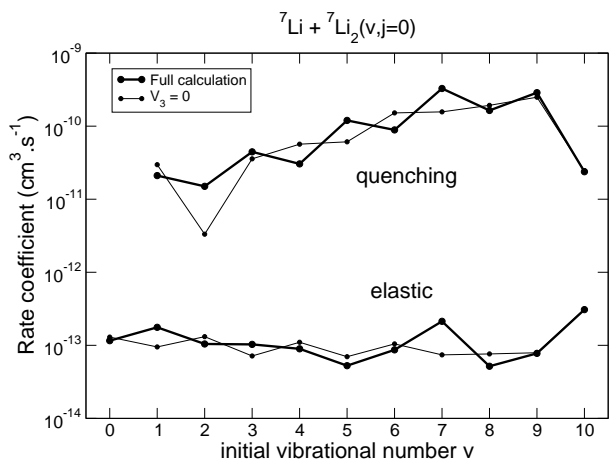}
\caption{Dependence of the elastic and quenching rate coefficients
on the vibrational excitation of the molecule
for $^7$Li + $^7$Li$_2(v=0-10,j=0)$ scattering at a collision energy of 10$^{-9}$~K.
The bold line corresponds to the full 
calculation and the thin line corresponds to the  
calculation without
the three-body term of the PES. 
Adapted  with permission from Qu\'em\'ener et al.~\cite{Quemener07}.
\label{RATEallv-LI3-FIG}
}
\end{center}
\end{figure}

Figure~\ref{RATEallv-LI3-FIG} shows the dependence of the rate coefficients for 
$^7$Li + $^7$Li$_2(v=0-10,j=0)$ collisions on the vibrational quantum number 
at a collision energy of 
10$^{-9}$~K~\cite{Quemener07}.
Quenching processes
are  more efficient than the elastic scattering for both high and low vibrational levels.
Similar results have been found for $^6$Li + $^6$Li$_2(v=0-9,j=1)$~\cite{Quemener07}
system composed of fermionic atoms.
Quenching rate coefficients show a slight decrease
when the molecule is in its highest
vibrational state.
This is because the overlap of the 
wavefunctions of highly excited diatomic molecules with low-lying 
vibrational levels is very small leading to small values for the interaction 
potential coupling matrix elements between the initial and final states~\cite{Stwalley04}.
These results 
do not directly apply to ultracold molecules composed
of fermionic atoms created 
near a Feshbach resonance. 
In these experiments, 
the quenching processes are suppressed 
because
the atom - atom scattering length
is tuned to large and positive values leading to 
efficient 
Pauli blocking mechanism,
as explained by Petrov et al.~\cite{Petrov04}.
In the theoretical study of Li + Li$_2$ collisions, 
the atom - atom scattering length is small and negative
and no suppression of the quenching processes
is found for the molecule in the last vibrational state.
Thus, the sign and magnitude of the Li$-$Li scattering length play 
a crucial role in
the mechanism  that suppresses quenching collisions.\\

The quenching rates displayed in 
Fig.~\ref{RATEallv-LI3-FIG} show an irregular dependence
on the vibrational state of the molecule.
This has already been seen for 
H + H$_2$ collisions~\cite{bala97a,Bodo06a}. 
In contrast, experimental measurements of quenching 
rate coefficients for Cs + Cs$_2$ collisions~\cite{Staanum06,Zahzam06}
do not show any dependence 
on the vibrational state of the molecule.
The differences between
these systems can be explained based on the following considerations.

First, the theoretical study applies to spin-polarized atom - dimer
alkali-metal systems whereas it is not the case for the experiment. 
A full theoretical treatment should involve the electronic and nuclear spins
of the alkali metal atoms as well as couplings between electronic surfaces
of different spins. 
This is beyond the scope of quantum dynamics calculations at present 
and will involve significant new code development and massive computational efforts.

Second, the dynamics of the two systems are different. 
The lithium system is lighter and has a more attractive three-body term 
than the cesium system~\cite{Soldan03}.
The triatomic adiabatic potential energy curves 
are well separated for a light system such as Li$_3$ 
and very dense
for a heavy system such as Cs$_3$.
The density of states
strongly influences the nature of vibrational  relaxation.
The effect of the density
of states has been illustrated in calculations for the Li + Li$_2$
system 
which exclude the three-body term. Removing the three-body term 
makes the energy levels more sparse.
This results in a more regular and monotonic
dependence of the rate coefficients on the vibrational levels $v=3-9$ as illustrated in Figure~\ref{RATEallv-LI3-FIG}.
This conclusion 
is in agreement with a previous work of 
Bodo et al.~\cite{Bodo06a}
for the H + H$_2$ system.
When they increased 
the density of states of the triatomic system,
they found no significant dependence on the vibrational states of the molecule.
However, for $v=10$, they obtained the same results with and without the
three-body term. 
The three-body term thus appears to be less significant
for collisions of molecules in high vibrational levels.
Since the three-body term vanishes at large separations 
it is only important in the short-range interaction region
while high vibrational states
involve spatially extended molecules 
and  sample the long range part of the interaction potential.
Therefore, three-body terms, which are the most difficult 
interaction energy terms to compute numerically for triatomic systems,
may be neglected to a first approximation in dynamics of highly vibrationally excited molecules.

In the experiments, rate coefficients have been measured for temperatures
of $40 \times 10^{-6}$~K~\cite{Zahzam06} and $60 \times 10^{-6}$~K~\cite{Staanum06} for Cs + Cs$_2$
while the theoretical rate coefficients for Li + Li$_2$
were reported for a temperature
of $10^{-9}$~K which corresponds to the Wigner threshold regime. Thus, it is likely that 
the experimental measurements did not probe the limiting values of the rate coefficients in the Wigner regime.

\subsubsection{Reactions of heteronuclear and isotopically substituted alkali-metal dimer systems}

Reactive collisions involving heteronuclear
molecules are currently of great interest.  A major goal of recent experiments has been
 to produce  heteronuclear alkali-metal dimers
in their electronic ground state. Examples include
RbCs~\cite{Kerman04,Sage05,Hudson08},
NaCs~\cite{Haimberger04,Kleinert07},
KRb~\cite{Mancini04,Wang04},
LiCs~\cite{Kraft06},
and mixed isotopes 
$^6$Li$^7$Li~\cite{Schloder01}.
Quantum dynamics calculations for heteronuclear systems
are more difficult. They are more interesting from a chemistry perspective
because the reactive channels in collisions of heteronuclear molecules 
can be distinguished from inelastic channels.\\

\begin{figure}[h]
\begin{center}
\begin{tabular}{cc}
\includegraphics*[height=8cm,keepaspectratio=false,angle=-90]{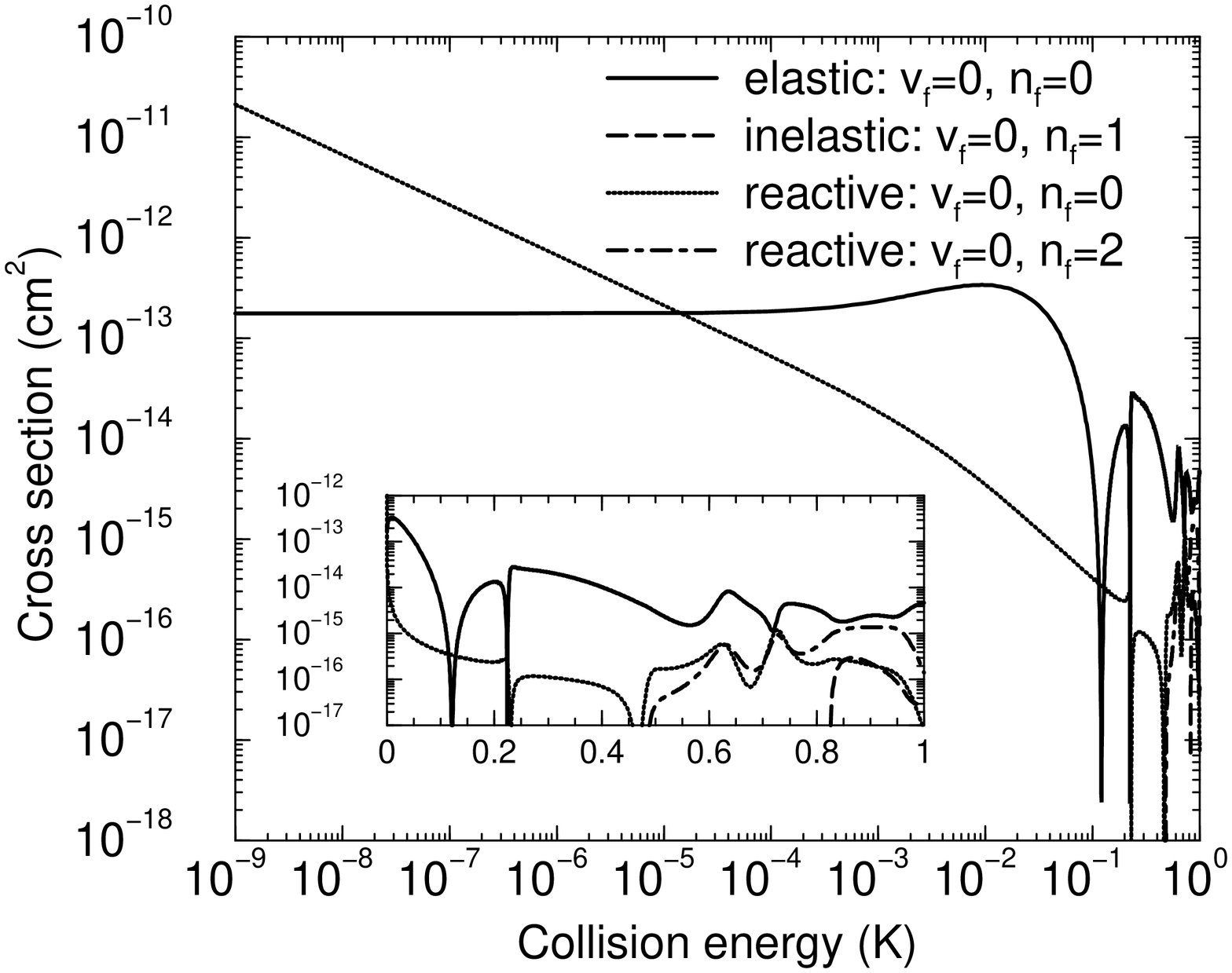}  \\
\includegraphics*[height=8cm,keepaspectratio=false,angle=-90]{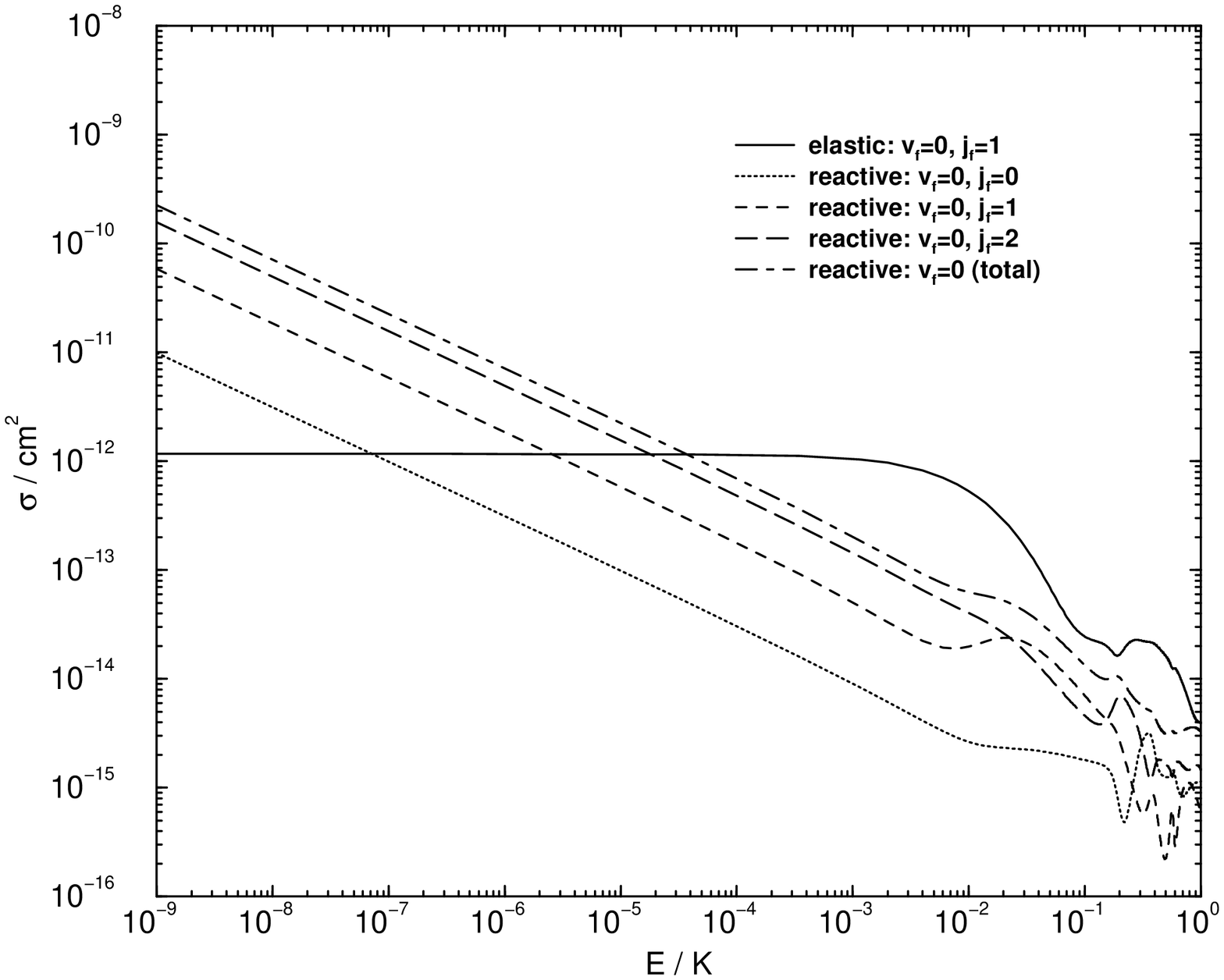}
\end{tabular}
\caption{Elastic and reactive s-wave cross sections for
$^7$Li + $^6$Li$^7$Li$(v=0,j=0)$ (upper panel) and
$^7$Li + $^6$Li$_2(v=0,j=1)$ (lower panel). Reproduced with permission 
from Cvita{\v s} et al.~\cite{Cvitas05b}.
\label{XSheteroLI3-FIG}
}
\end{center}
\end{figure}

Cvita{\v s} et al.~\cite{Cvitas05b,Cvitas07} have explored the quantum dynamics of
 $^7$Li + $^6$Li$^7$Li$(v=0,j=0)$, $^7$Li + $^6$Li$_2(v=0,j=1)$,
$^6$Li + $^7$Li$_2(v=0,j=0)$, and $^6$Li + $^6$Li$^7$Li$(v=0,j=0)$ collisions.
The $J=0$ cross sections for elastic scattering and chemical reactions in collisions of 
 $^7$Li with $^6$Li$^7$Li$(v=0,j=0)$
are presented in the upper panel in Fig.~\ref{XSheteroLI3-FIG} for a wide range of
 collision energies.
The cross sections are believed to be converged for energies up to 10$^{-4}$~K.
The reactive process which leads to
$^6$Li + $^7$Li$_2(v=0,j=0)$ dominates over the elastic scattering
at ultralow energies.
However, the reactive process 
is less efficient than the vibrational relaxation
process
in homonuclear alkali systems
discussed above. 
For instance, at a collision energy of 10$^{-9}$~K,
the ratio of the cross sections for reactive and elastic collisions
is about 
two orders of magnitude
for the heteronuclear system presented in Fig.~\ref{XSheteroLI3-FIG}
whereas it is 
about three orders of magnitude for homonuclear systems.
Cvita{\v s} et al. attributed the
smaller ratio
to the presence of only  one 
open channel in the heteronuclear reaction.

The $J=1$ elastic and reactive cross sections for $^7$Li + $^6$Li$_2(v=0,j=1)$ collisions
are presented in the lower panel of Fig.~\ref{XSheteroLI3-FIG}.
The reactive process  leading to the formation of $^6$Li$^7$Li molecules
is slightly more efficient than elastic scattering in the Wigner regime.
The other possible collision processes are $^6$Li + $^6$Li$^7$Li$(v=0,j=0)$
and $^6$Li + $^7$Li$_2$$(v=0,j=0)$. However, only elastic scattering occurs in these
systems at ultralow collisions energies.

These results provide important implications for the experiments on the production of 
$^6$Li$^7$Li in an ultracold mixture of $^6$Li and $^7$Li atoms.
Cvita{\v s} et al. proposed  to remove quickly the $^7$Li
atomic gas after the formation of the $^6$Li$^7$Li molecules in their ground state
to prevent the destructive reactive processes. 
But keeping
the $^6$Li atomic gas is recommended for  
sympathetic cooling of the $^6$Li$^7$Li molecules
since only elastic collisions are possible.
Removing  the  $^6$Li atoms
will leave the fermionic heteronuclear $^6$Li$^7$Li molecules 
in the trap.
In the absence of $^6$Li atoms, 
evaporative cooling by collisions between fermionic $^6$Li$^7$Li molecules 
will not be effective since s-wave collisions of identical 
fermionic dimers will be suppressed due to the Pauli exclusion principle. \\

Table~\ref{TAB4} provides a compilation of zero-temperature quenching rate coefficients for 
different alkali-metal trimer systems.

\begin{table}[h]
\begin{center}
\begin{tabular}{c c c c}
\hline
system & initial $(v,j)$ & $k_{T=0}$ (cm$^3$s$^{-1}$) & Ref.  \\ [0.5ex]
\hline
$^{39}$K + $^{39}$K$_2$ & $(v=1,j=0)$ & 1.1 $\times$ 10$^{-10}$ & \cite{Quemener05} \\
$^{40}$K + $^{40}$K$_2$ & $(v=1,j=1)$ & 8.0 $\times$ 10$^{-11}$ & \cite{Quemener05} \\
$^{41}$K + $^{41}$K$_2$ & $(v=1,j=0)$ & 9.8 $\times$ 10$^{-11}$ & \cite{Quemener05} \\
\hline
$^7$Li + $^7$Li$_2$ & $(v=1,j=0)$ & 2.1 $\times$ 10$^{-11}$ & \cite{Quemener07} \\
                    & $(v=2,j=0)$ & 1.5 $\times$ 10$^{-11}$ & \cite{Quemener07} \\
                    & $(v=3,j=0)$ & 4.4 $\times$ 10$^{-11}$ & \cite{Quemener07} \\
                    & $(v=4,j=0)$ & 3.0 $\times$ 10$^{-11}$ & \cite{Quemener07} \\
                    & $(v=5,j=0)$ & 1.2 $\times$ 10$^{-10}$ & \cite{Quemener07} \\
                    & $(v=6,j=0)$ & 8.9 $\times$ 10$^{-11}$ & \cite{Quemener07} \\
                    & $(v=7,j=0)$ & 3.3 $\times$ 10$^{-10}$ & \cite{Quemener07} \\
                    & $(v=8,j=0)$ & 1.6 $\times$ 10$^{-10}$ & \cite{Quemener07} \\
                    & $(v=9,j=0)$ & 2.9 $\times$ 10$^{-10}$ & \cite{Quemener07} \\
                    & $(v=10,j=0)$ & 2.4 $\times$ 10$^{-11}$ & \cite{Quemener07} \\
$^6$Li + $^6$Li$_2$ & $(v=1,j=1)$ & 3.3 $\times$ 10$^{-11}$ & \cite{Quemener07} \\
                    & $(v=2,j=1)$ & 2.0 $\times$ 10$^{-11}$ & \cite{Quemener07} \\
                    & $(v=3,j=1)$ & 5.1 $\times$ 10$^{-11}$ & \cite{Quemener07} \\
\hline
$^7$Li + $^7$Li$_2$ & $(v=1,j=0)$ & 5.6 $\times$ 10$^{-10}$ & \cite{Cvitas05a} \\
                    & $(v=2,j=0)$ & 9 $\times$ 10$^{-11}$ & \cite{Cvitas05a} \\
$^6$Li + $^6$Li$_2$ & $(v=1,j=1)$ & 2.8 $\times$ 10$^{-10}$ & \cite{Cvitas05a} \\
                    & $(v=2,j=1)$ & 4 $\times$ 10$^{-10}$ & \cite{Cvitas05a} \\
\hline
$^{23}$Na + $^{23}$Na$_2$ & $(v=1,j=0)$ & 2.9 $\times$ 10$^{-10}$ & \cite{Quemener04} \\
                          & $(v=2,j=0)$ & 1.1 $\times$ 10$^{-10}$ & \cite{Quemener04} \\
                          & $(v=3,j=0)$ & 6.1 $\times$ 10$^{-11}$ & \cite{Quemener04} \\
\hline
$^7$Li + $^6$Li$^7$Li & $(v=0,j=0)$ & 4.1 $\times$ 10$^{-12}$ & \cite{Cvitas07} \\
                      & $(v=1,j=0)$ & 2.1 $\times$ 10$^{-10}$ & \cite{Cvitas07} \\
                      & $(v=2,j=0)$ & 4.4 $\times$ 10$^{-10}$ & \cite{Cvitas07} \\
                      & $(v=3,j=0)$ & 4.0 $\times$ 10$^{-10}$ & \cite{Cvitas07} \\
$^7$Li + $^6$Li$_2$ & $(v=0,j=1)$ & 4.4 $\times$ 10$^{-11}$ & \cite{Cvitas07} \\
                    & $(v=1,j=1)$ & 5.2 $\times$ 10$^{-10}$ & \cite{Cvitas07} \\
                    & $(v=2,j=1)$ & 2.6 $\times$ 10$^{-10}$ & \cite{Cvitas07} \\
                    & $(v=3,j=1)$ & 3.0 $\times$ 10$^{-10}$ & \cite{Cvitas07} \\
$^6$Li + $^6$Li$^7$Li & $(v=1,j=0)$ & 2.6 $\times$ 10$^{-10}$ & \cite{Cvitas07} \\
                      & $(v=2,j=0)$ & 3.5 $\times$ 10$^{-10}$ & \cite{Cvitas07} \\
                      & $(v=3,j=0)$ & 4.4 $\times$ 10$^{-10}$ & \cite{Cvitas07} \\
$^6$Li + $^7$Li$_2$ & $(v=1,j=0)$ & 2.8 $\times$ 10$^{-10}$ & \cite{Cvitas07} \\
                    & $(v=2,j=0)$ & 5.3 $\times$ 10$^{-10}$ & \cite{Cvitas07} \\
                    & $(v=3,j=0)$ & 4.6 $\times$ 10$^{-10}$ & \cite{Cvitas07} \\ [1ex]
\hline
\end{tabular}
\end{center}
\caption{Zero-temperature quenching rate coefficients for different atom - dimer 
alkali  metal systems.
\label{TAB4}
}
\end{table}

\section{ Inelastic molecule - molecule collisions}

Many of the studies of cold and ultracold molecules have focused on reactive and non-reactive scattering in
atom - molecule collisions. At high densities of trapped molecules, molecule - molecule collisions need to be considered. The presence of rotational and vibrational degrees of freedom in both collision partners 
 make molecule - molecule systems especially interesting. 
However, quantum dynamics calculations of molecule - molecule
collisions are significantly more challenging. Most dynamical
 calculations of molecule - molecule scattering have relied on
 the rigid rotor approximation and some
recent studies have adopted the coupled-states approximation. Calculations performed at high
collision energies have employed more approximate methods based on semi-classical techniques. Here, we present
a brief account of recent dynamical calculations for
 the H$_2$ + H$_2$ system and discuss some ongoing work on
full-dimensional quantum calculations of ro-vibrational 
transitions in H$_2-$H$_2$ collisions. A brief discussion 
of hyperfine transitions in molecule - molecule systems is also provided 
for the illustrative examples of the O$_2$ + O$_2$ and OH + OH / OD + OD systems.

\subsection{ Molecules in the ground vibrational state}

The H$_2$ + H$_2$ system
is the simplest neutral tetra-atomic system and it 
serves
as a prototype for describing collisions 
between diatomic molecules. 
Though there have been a number of experimental and theoretical studies 
over the past several years on the H$_2$ + H$_2$ system (see Ref.~\cite{Teck-Lee06} and references therein),
only a few studies have explored the collision dynamics in the cold and ultracold regime. \\

\begin{figure}[h]
\begin{center}
\includegraphics*[width=8cm,keepaspectratio=true]{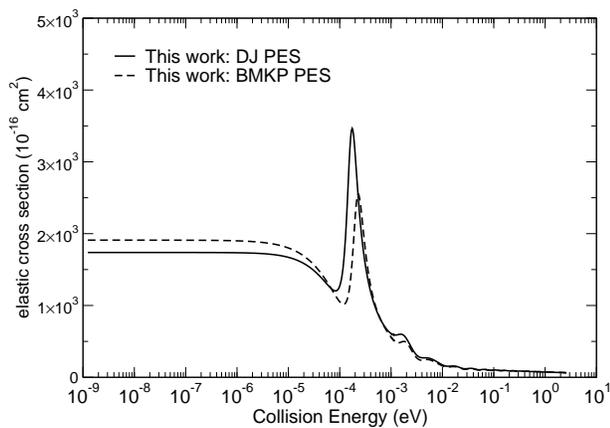}
\caption{Elastic cross section of H$_2(v=0,j=0)$ + H$_2(v'=0,j'=0)$ as a function
of the collision energy. 
Reproduced  with permission from Lee et al.~\cite{Teck-Lee06}.
\label{XSEL-H2H2vj0000-FIG}
}
\end{center}
\end{figure}

Forrey~\cite{Forrey01} presented a study of 
 rotational transitions in H$_2(v=0,j=2)$ + H$_2(v'=0,j'=2)$ 
collisions at cold and ultracold collision energies using a rigid rotor model. 
He calculated the real and imaginary components of the complex scattering length 
for H$_2(v=0,j=2,4,6,8)$ + H$_2(v'=0,j'=j)$ collisions
and showed that the imaginary parts decrease with increasing $j,j'$. 
The imaginary parts of the scattering lengths are found to be
small compared to the real parts, leading to small inelastic cross sections. 
Mat\'e et al.~\cite{Mate05} reported an experimental study 
of the rate coefficient for the
H$_2(v=0,j=0)$ + H$_2(v'=0,j'=0)$ $\to$ 
H$_2(v_f=0,j_f=0)$ + H$_2(v'_f=0,j'_f=2)$ transition
at temperatures between 2~K and 110~K.
They found good agreement between the experimental results 
and quantum dynamics calculations based on the rigid rotor model
using a PES reported by Diep and Johnson (DJ)~\cite{Diep00}. 
Montero et al.~\cite{Montero06} have investigated cold inelastic collisions of
n-H$_2$ molecules in the ground vibrational state, using a 3:1 gas mixture of ortho and para hydrogen.
They obtained good agreement between the experimental data and the
 theoretical calculations based on the DJ PES.\\

Using a quantum formalism based on the rigid rotor model, Lee et al.~\cite{Teck-Lee06}
have recently presented a comparative analysis 
of cross sections for rotationally-inelastic collisions between H$_2$ molecules
at low and ultra-low energies. 
The elastic cross sections for H$_2(v=0,j=0)$ + H$_2(v'=0,j'=0)$ collisions 
obtained in this study
are presented in Fig.~\ref{XSEL-H2H2vj0000-FIG} for two different PESs.
The limiting value of the elastic scattering cross section in the 
ultra-low energy regime is $1.91 \times 10^{-13}$~cm$^2$
with the PES of Boothroyd, Martin, Keogh and Peterson (BMKP)~\cite{Boothroyd02} 
and $1.74 \times 10^{-13}$~cm$^2$  
with the DJ PES~\cite{Diep00}. 
For low collision energies the dynamics 
is sensitive to higher-order anisotropic terms in the angular expansion of the interaction potential. 
A diatom - diatom scattering length
of 5.88~\AA~was obtained for the DJ PES and 6.16~\AA~for the BMKP PES. 
Cross sections for the quenching of rotationally excited H$_2$ molecules in H$_2(v=0,j=2)$ + H$_2(v'=0,j'=0)$
and H$_2(v=0,j=2)$ + H$_2(v'=0,j'=2)$ collisions were presented  at low and ultra-low energies. \\

Quantum calculations of rotational relaxation of CO in cold and ultracold
 collisions with H$_2$ have recently been performed by 
Yang et al. ~\cite{yang06a,yang06b}.
They reported
quenching rate coefficients for $j=1-3$ of the CO molecule in collisions with both ortho- and para-H$_2$ ~\cite{yang06a}.
 Due to the relatively deep van der Waals interaction potential for the H$_2-$CO system the cross sections
exhibit a number of narrow resonances for collision energies between $1.0-40.0$~cm$^{-1}$. The signatures of
these resonances are present in the temperature dependence of the rate coefficient which shows broad oscillatory
features in the temperature range of $10^{-2} - 50$~K ~\cite{yang06a}. \\

Bohn and co-workers have reported extensive calculations of hyperfine transitions in ultracold molecule - molecule collisions. 
Avdeenkov and Bohn studied ultracold collisions between O$_2$ molecules~\cite{Avdeenkov01}. 
They used a quantum mechanical formalism based on the rigid rotor model and included
the electronic spin structure of the O$_2$ molecules. Couplings between the 
rotational angular momentum and the electronic spin of the molecules
lead to rotational fine structure.
Avdeenkov and Bohn discussed the elastic and inelastic loss processes in
$^{17}$O$_2$ + $^{17}$O$_2$ and  $^{16}$O$_2$ + $^{16}$O$_2$ collisions.
They found that for collision energies below 10$^{-2}$~K, elastic collision
cross sections
are larger than inelastic spin-flipping transitions. 
Based on relative magnitudes of elastic and inelastic spin-flipping cross sections they
concluded that $^{17}$O$_2$ molecules 
would be good candidates for evaporative cooling.
In contrast, for $^{16}$O$_2$ + $^{16}$O$_2$ collisions,
inelastic processes are more efficient than elastic collisions  so that 
$^{16}$O$_2$ molecules are prone to collisional trap loss.

\begin{figure}[h]
\begin{center}
\includegraphics*[bb=90 0 590 792,height=9.0cm,keepaspectratio=true,angle=-90]{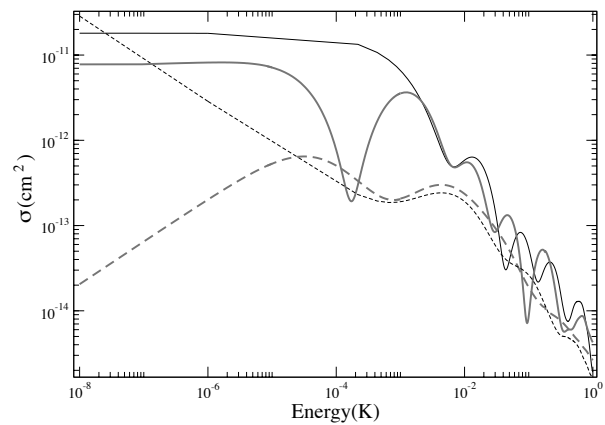}  
\caption{
Elastic and inelastic cross sections for OD + OD (thick gray line) 
and OH + OH (thin black line) collisions
for an applied electric field
of $\varepsilon$=100~V/cm.
Solid and dashed lines
refer to elastic and inelastic cross sections, respectively.
Reproduced with permission from Avdeenkov et al.~\cite{Avdeenkov05}.
\label{RATE-OHOD-FIG}
}
\end{center}
\end{figure}

Avdeenkov and Bohn also studied ultracold collisions 
between OH~\cite{Avdeenkov02,Avdeenkov03} 
and OD radicals~\cite{Avdeenkov05} in the presence of
 an applied electric field.
They showed that~\cite{Avdeenkov05},
elastic scattering is more efficient
than inelastic processes for ultracold collisions between fermionic OD molecules,
inhibiting state changing collisions. 
The  energy dependence of elastic and inelastic cross sections
for OH + OH and OD + OD collisions is illustrated in Fig.~\ref{RATE-OHOD-FIG}
for an applied electric field
of $\varepsilon$=100~V/cm.
While the elastic cross sections approach finite values in the Wigner regime for the
bosonic system of OH molecules
and the fermionic system of OD molecules, the inelastic cross sections
exhibit a totally different behavior.
At ultra-low energies when molecules interact with an electric field,
s-wave scattering 
yields an 
inelastic cross section diverging as E$_{\text coll}^{-1/2}$ for bosonic systems,
while p-wave scattering
yields an 
inelastic cross section vanishing as E$_{\text coll}^{1/2}$
for fermionic systems.
Thus, the inelastic processes are suppressed
for the fermionic system.
The  differences are attributed to 
the bosonic/fermionic character of the molecules
and to the applied electric field.
In the absence of an applied electric field, the elastic cross section
of the fermionic system would decrease as E$_{\text coll}^2$,
faster than the inelastic cross section, which
decreases as E$_{\text coll}^{1/2}$.
Ticknor and Bohn~\cite{Ticknor05} subsequently studied
OH$-$OH  collisions
in the presence of a magnetic field. 
They showed that magnetic fields of several thousand gauss
reduce inelastic collisions
by about two orders of magnitude.
Based on these results, they concluded that
magnetic trapping
may be favorable for OH molecules.

\subsection{ Vibrationally inelastic transitions}

The theoretical studies presented above for molecule - molecule collisions refer to
rigid rotor molecules. 
Pogrebnya and Clary~\cite{Pogrebnya02a,Pogrebnya02b} have  investigated
vibrational relaxation in collisions of 
hydrogen molecules using a full-dimensional quantum dynamics formalism
with an angular momentum decoupling approximation in the body fixed frame.
They studied H$_2(v=1,j)$ + H$_2(v=0,j')$ collisions involving
both para-para and ortho-ortho combinations
for collision energies
of 1~meV (11.6~K) to 1~eV (11604~K). Time-dependent quantum mechanical calculations based on the 
multiconfiguration time-dependent Hartree approach have been recently applied to study
 full-dimensional
 quantum dynamics of the H$_2-$H$_2$ system~\cite{Panda07,Otto08}. These
 methods are suitable at high collision 
energies and have not been applied to cold and ultracold collisions.
Very recently, vibrational energy transfer mechanism for ultracold collisions between 
para-hydrogen molecules  
has been explored by Qu{\'e}m{\'e}ner et al.~\cite{Quemener08} 
using a full-dimensional quantum theoretical formalism
implemented in a new code
written by Krems~\cite{Roman06}. The quantum scattering calculations are based on the theory
described by
Arthurs and Dalgarno~\cite{Arthurs60},
Takayanagi~\cite{Takayanagi65},
Green~\cite{Green75},
and Alexander and DePristo~\cite{Alexander77}. 
The H$_2$ molecules were initially in different quantum states 
characterized by the vibrational quantum number $v$ and the rotational angular momentum $j$. 
A combination of two ro-vibrational states of H$_2$ was referred to as a combined molecular state (CMS). 
A CMS, denoted as $(v j v' j')$, represents a unique quantum state of 
the diatom - diatom system
before or after a collision.

\begin{figure}[h]
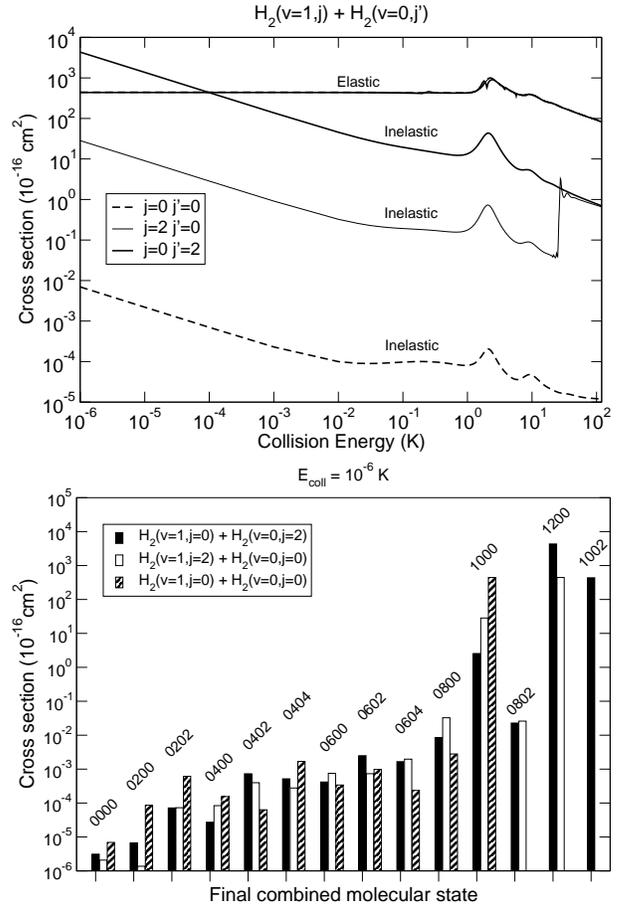

\begin{center}
\begin{tabular}{cc}
\includegraphics*[width=8cm,keepaspectratio=true]{Figure25a.eps} \\
\includegraphics*[width=8cm,keepaspectratio=true]{Figure25b.eps}
\end{tabular}
\caption{Elastic and inelastic cross sections for the collisions
H$_2(v=1,j=0)$ + H$_2(v=0,j=0)$, H$_2(v=1,j=2)$ + H$_2(v=0,j=0)$, and H$_2(v=1,j=0)$ + H$_2(v=0,j=2)$:
total cross sections (upper panel); state-to-state cross sections at 10$^{-6}$~K (lower panel).
Adapted  with permission from Qu\'em\'ener et al.~\cite{Quemener08}.
\label{XSELIN-H2H2-FIG}
}
\end{center}
\end{figure}

The cross sections for H$_2(v=1,j=0)$ + H$_2(v=0,j=0)$, H$_2(v=1,j=2)$ + H$_2(v=0,j=0)$ 
and H$_2(v=1,j=0)$ + H$_2(v=0,j=2)$
collisions are presented in the upper panel of Fig.~\ref{XSELIN-H2H2-FIG}.
Using the CMS notation, this corresponds, respectively, to initial CMSs (1000), (1200) and (1002).
The elastic cross sections are almost independent of the different initial ro-vibrational states of the molecules,
but the inelastic  cross sections are strongly dependent on the initial rotational and vibrational levels of the H$_2$ molecules. 
At 10$^{-6}$~K, the inelastic relaxation of H$_2(v=1,j=0)$ is almost six
orders of magnitude more efficient in collisions of H$_2(v=0,j=2)$ than in collisions with H$_2(v=0,j=0)$
and two orders of magnitude more efficient than in collisions of H$_2(v=1,j=2)$ with H$_2(v=0,j=0)$.
At 25.45~K, which corresponds to the energy difference between 
the CMSs (1002) and (1200),
the inelastic cross sections become comparable
for  H$_2(v=1,j=2)$ + H$_2(v=0,j=0)$ and H$_2(v=1,j=0)$ + H$_2(v=0,j=2)$ collisions.
The inelastic scattering depends on the type and the combination of ro-vibrational levels involved in the collision,
whether it involves  ground state molecules,
H$_2(v=0,j=0)$, vibrationally excited molecules, H$_2(v=1,j=0)$, rotationally excited molecules H$_2(v=0,j=2)$, 
or ro-vibrationally excited molecules, H$_2(v=1,j=2)$. 
H$_2$ molecules are weakly interacting and characterized by shallow 
van der Waals interaction at large separations and the computed rate coefficients are not representative 
of strongly interacting alkali-metal dimer systems. 
For example, 
Mukaiyama et al.~\cite{Mukaiyama04} reported an inelastic rate coefficient of
$5.1 \times 10^{-11}$~cm$^3$~s$^{-1}$ 
for collisions between two weakly bound Na$_2$ molecules created by the Feshbach resonance method. 
In a similar study Syassen et al.~\cite{Syassen06} reported a rate coefficient of $3 \times 10^{-10}$~cm$^3$~s$^{-1}$
for collisions between two Rb$_2$ molecules.
Zahzam et al.~\cite{Zahzam06} estimated 
a rate coefficient of about $10^{-11}$~cm$^3$~s$^{-1}$
for collisions between Cs$_2$ molecules.
Ferlaino et al.~\cite{Ferlaino08} measured a rate coefficient of $9 \times 10^{-11}$~cm$^3$~s$^{-1}$
for collisions between Cs$_2$ molecules in the non-halo regime. The latter authors also measured 
the variation of the inelastic rate coefficient as a function of the 
atom - atom scattering length
for collisions between tunable halo dimers of Cs$_2$.
The large values of the inelastic rate coefficients for
alkali-metal dimer systems are attributed to
strong inelastic couplings and deeper potential energy wells.

The state-to-state cross sections for three 
ro-vibrational combinations of the H$_2-$H$_2$ system are presented in the lower panel 
of Fig.~\ref{XSELIN-H2H2-FIG}.
The magnitude of the inelastic cross sections
depends on the propensity of the diatom - diatom system to conserve
the internal energy and the total rotational angular momentum of the colliding molecules~\cite{Quemener08}.
The final 
state-to-state distribution in H$_2(v=1,j=0)$ + H$_2(v=0,j=0)$ collisions shows that 
there is no preferential population of rotational quantum number of either of the colliding molecules.
The conservation of the total rotational angular momentum would entail a large change in the internal 
energy of the molecules. Thus, the purely vibrational transition (1000) $\to$ (0000) 
is not efficient because the energy gap is large.
On the other hand, (near) conservation of the internal energy requires a
large change in the total rotational angular momentum of the colliding 
molecules and the transition (1000) $\to$ (0800) is not dominant either.
However, the    
state-to-state cross sections for H$_2(v=1,j=2)$ + H$_2(v=0,j=0)$ collisions indicate
that
the transition (1200) $\to$ (1000)
is more efficient than all the other transitions combined.
For this transition, the total rotational angular momentum change
and the internal energy transfer are both minimized,
leading to a more efficient energy transfer process.
The state-to-state cross sections for H$_2(v=1,j=0)$ + H$_2(v=0,j=2)$ 
collisions present an interesting scenario in which 
the transition (1002) $\to$ (1200) is highly efficient and selective.
In this case, the total rotational angular momentum is conserved
and the internal energy is almost unchanged 
(the energy gap between (1200) and (1002) is only 25.45~K).
This creates a favorable situation
leading to a near-resonant energy transfer process.
This particular mechanism cannot occur in atom - diatom 
systems because simultaneous conservation of rotational angular momentum
and internal energy of the molecule cannot occur.
The mechanism
is reminiscent of quasi-resonant energy transfer in collisions of
rotationally excited diatomic molecules with atoms discussed previously~\cite{forrey99a,stewart88,magill88,Heller06},
but with a purely quantum origin.
The near-resonant process may be an important mechanism for 
collisional energy
transfer in ultracold molecules formed by photoassociation of ultracold
atoms and for chemical reactions producing identical
molecules. \\

Table~\ref{TAB5} provides a compilation of zero-temperature rate coefficients for 
rotational and vibrational quenching in molecule - molecule systems.

\begin{table}[h]
\begin{center}
\begin{tabular}{c c c c}
\hline
system & initial $(v,j,v',j')$ & $k_{T=0}$ (cm$^3$s$^{-1}$) & Ref.  \\ [0.5ex]
\hline
H$_2$ + H$_2$ & $(v=1,j=0,v'=0,j'=0)$ & 8.9 $\times$ 10$^{-18}$ & \cite{Quemener08} \\
              & $(v=2,j=0,v'=0,j'=0)$ & 3.9 $\times$ 10$^{-17}$ & \cite{Quemener08} \\
              & $(v=3,j=0,v'=0,j'=0)$ & 1.2 $\times$ 10$^{-16}$ & \cite{Quemener08} \\
              & $(v=4,j=0,v'=0,j'=0)$ & 1.8 $\times$ 10$^{-16}$ & \cite{Quemener08} \\
              & $(v=5,j=0,v'=0,j'=0)$ & 6.9 $\times$ 10$^{-16}$ & \cite{Quemener08} \\
              & $(v=6,j=0,v'=0,j'=0)$ & 2.9 $\times$ 10$^{-15}$ & \cite{Quemener08} \\
              & $(v=1,j=0,v'=1,j'=0)$ & 2.7 $\times$ 10$^{-16}$ & \cite{Quemener08} \\
              & $(v=2,j=0,v'=1,j'=0)$ & 1.3 $\times$ 10$^{-14}$ & \cite{Quemener08} \\
              & $(v=3,j=0,v'=1,j'=0)$ & 1.6 $\times$ 10$^{-14}$ & \cite{Quemener08} \\
              & $(v=4,j=0,v'=1,j'=0)$ & 9.0 $\times$ 10$^{-15}$ & \cite{Quemener08} \\
              & $(v=2,j=0,v'=2,j'=0)$ & 6.1 $\times$ 10$^{-16}$ & \cite{Quemener08} \\
\hline
H$_2$ + H$_2$ & $(v=0,j=2,v'=0,j'=0)$ & 3.9 $\times$ 10$^{-14}$ & \cite{Quemener08} \\
              & $(v=1,j=2,v'=0,j'=0)$ & 3.7 $\times$ 10$^{-14}$ & \cite{Quemener08} \\
              & $(v=2,j=2,v'=0,j'=0)$ & 8.7 $\times$ 10$^{-14}$ & \cite{Quemener08} \\
              & $(v=3,j=2,v'=0,j'=0)$ & 1.5 $\times$ 10$^{-13}$ & \cite{Quemener08} \\
              & $(v=4,j=2,v'=0,j'=0)$ & 2.5 $\times$ 10$^{-13}$ & \cite{Quemener08} \\
              & $(v=1,j=2,v'=1,j'=0)$ & 8.8 $\times$ 10$^{-14}$ & \cite{Quemener08} \\
              & $(v=2,j=2,v'=1,j'=0)$ & 6.1 $\times$ 10$^{-14}$ & \cite{Quemener08} \\
\hline
H$_2$ + H$_2$ & $(v=1,j=0,v'=0,j'=2)$ & 5.6 $\times$ 10$^{-12}$ & \cite{Quemener08} \\
              & $(v=2,j=0,v'=0,j'=2)$ & 3.4 $\times$ 10$^{-12}$ & \cite{Quemener08} \\
              & $(v=3,j=0,v'=0,j'=2)$ & 2.7 $\times$ 10$^{-12}$ & \cite{Quemener08} \\
              & $(v=4,j=0,v'=0,j'=2)$ & 2.1 $\times$ 10$^{-12}$ & \cite{Quemener08} \\
              & $(v=2,j=0,v'=1,j'=2)$ & 5.2 $\times$ 10$^{-12}$ & \cite{Quemener08} \\ 
\hline
H$_2$ + CO & $(v=0,j=0,v'=0,j'=1)$ & 2.0 $\times$ 10$^{-12}$ & \cite{yang06a} \\
           & $(v=0,j=0,v'=0,j'=2)$ & 3.0 $\times$ 10$^{-11}$ & \cite{yang06a} \\
           & $(v=0,j=0,v'=0,j'=3)$ & 1.2 $\times$ 10$^{-10}$ & \cite{yang06a} \\
\hline
H$_2$ + CO & $(v=0,j=1,v'=0,j'=1)$ & 1.2 $\times$ 10$^{-11}$ & \cite{yang06a} \\
           & $(v=0,j=1,v'=0,j'=2)$ & 4.0 $\times$ 10$^{-11}$ & \cite{yang06a} \\
           & $(v=0,j=1,v'=0,j'=3)$ & 8.5 $\times$ 10$^{-11}$ & \cite{yang06a} \\ [1ex]
\hline
\end{tabular}
\end{center}
\caption{Zero-temperature inelastic rate coefficients for different molecule - molecule systems.
\label{TAB5}
}
\end{table}

\section{ Summary and outlook}

In this chapter we have given an overview of recent theoretical
 studies of atom - molecule and molecule - molecule 
collisions at cold and ultracold temperatures. Though such systems have been extensively studied at higher 
collision energies over the last few decades, the new experimental breakthroughs in creating dense samples
 of cold and ultracold molecules have provided unprecedented opportunities to explore
elastic, inelastic, and reactive collisions at temperatures close to absolute zero. These studies have
 revealed unique aspects of molecular collisions and energy transfer mechanisms that are otherwise not
 evident in thermal energy collisions. \\

The long duration of collisions combined with large de Broglie wavelengths at cold and ultracold 
temperatures leads to interesting quantum effects. Calculations have shown that reactions with 
insurmountable energy barriers may still occur at temperatures close to absolute zero, and in certain
 cases, with appreciable rate coefficients. Such tunneling dominated reactions have been the topic of many 
recent investigations and may soon be amenable to experimental investigation in the cold and ultracold 
regime. That the rates of these reactions can be enhanced by vibrational excitation of the 
molecule is an interesting scenario for experimental studies of ultracold chemical reactions. \\

The studies of alkali-metal homonuclear and heteronuclear trimer systems have 
stimulated considerable
experimental interest in investigating  chemical reactivity at ultracold temperatures. The 
challenge for theory is to describe collisions involving highly vibrationally excited 
molecules. Recent theoretical calculations have indicated that the three-body interaction potential can be 
neglected for highly vibrationally excited molecules offering significant savings in computational effort.
Even so, heavier alkali-metal trimer systems pose a daunting computational challenge. \\

The quantitative description of ultracold molecule - molecule collisions is another challenging topic. The 
recent progress on the H$_2-$H$_2$ system will be difficult to implement for heavier systems due to the large
number of ro-vibrational levels of the molecules. The study of H$_2-$H$_2$ collisions
 has shown that, for certain 
combinations of ro-vibrational levels, the energy transfer may occur to specific final ro-vibrational
states. In such cases, the calculations can use a much smaller basis set without compromising the
accuracy. \\

Currently there is substantial interest in controlling the collisional outcome using external electric and
magnetic fields. While the idea of coherent control of molecular collisions and chemical reactivity has 
existed for a long time and some important progress has been achieved, the possibility of creating 
coherent and dense samples of molecules in specific quantum states has given further impetus to the field 
of controlled chemistry. We expect that the coming years will see a far greater activity in this direction
driven by cold and ultracold molecules and also by the possibility of controlling chemical reactivity 
using external electric and magnetic fields. Electronically non-adiabatic effects in ultracold collisions
is a largely unexplored area, which can be expected to attract much attention. \\

Acknowledgments: This work was supported by NSF grants \# PHY-0555565 (N.B.), AST-0607524 (N.B.),
and by the Chemical Science, Geoscience and Bioscience Division of the Office
of Basic Energy Science, Office of Science, U.S. Department of Energy (A.D.).

\end{document}